\documentclass[graphicx,reprint,twocolumn,nofootinbib, superscriptaddress,aip,pop]{revtex4-1}
\usepackage{graphics,bm,overpic,subfigure,color,amsmath,multirow,hyperref,url}

\newcommand{\pd}[3][]{\ensuremath{ \frac{\partial^{#1} #2} {\partial #3} } }
\newcommand{\gyror}[1]{\ensuremath{ {\left< #1 \right>}_{\bm{r}}}}

\newcommand{\gyroR}[2][s]{\ensuremath{{\left< #2 \right>}_{\bm{R}_{#1}}}}

\newcommand{\uv}[1]{\bm{\hat{#1}}} 

\newcommand{\Oi}{\Omega_{i}}

\newcommand{\fsr}{\gamma_{E}}
\newcommand{\tprim}{\kappa}

\newcommand{\gstwolnorm}{R_0} 
\newcommand{\gstworefspec}{i} 

\newcommand{\gstwofsrnorm}{} 
\newcommand{\gstwoQnormflatbd}{n_\gstworefspec T_\gstworefspec v_{th\gstworefspec} \rho_{\gstworefspec}^2 / 2 \sqrt{2} \gstwolnorm^2 } 

\newcommand{\pa}{_{\parallel}}
\newcommand{\pp}{_{\perp}}

\newcommand{\oov}[1]{\frac{1}{#1}}

\newcommand{\vct}[1]{\bm{#1}}

\newcommand{\pdt}[1]{\pd{#1}{t}} 
\newcommand{\dbd}[2]{\frac{d#1}{d#2}} 
\newcommand{\ddt}[1]{\dbd{#1}{t}} 

\newcommand{\lp}{\left(}
\newcommand{\rp}{\right)}

\newcommand{\lsq}{\left[}
\newcommand{\rsq}{\right]}

\newcommand{\md}[1]{\left| #1 \right|}

\newcommand{\btor}{B_{\phi}}
\newcommand{\bpol}{B_{\theta}}

\newcommand{\rudolphpeierls}{Rudolph Peierls Centre for Theoretical Physics, University of Oxford, 1 Keble Road,  Oxford, OX1 3NP, UK}
\newcommand{\blackett}{Blackett Laboratory, Imperial College, London, SW7 2AZ, UK}
\newcommand{\culham}{EURATOM/CCFE Fusion Association, Culham Science Centre, Abingdon, OX14 3DB, UK}

\newcommand{\wpauli}{Wolfgang Pauli Institute, University of Vienna, Nordbergstrasse 15, A 1090 Wien}

\newcommand{\edmundhighcock}{
\author{E.\ G.\ Highcock}
\email{edmund.highcock@physics.ox.ac.uk}
\affiliation{
\rudolphpeierls
}
\affiliation{
\culham
}
\affiliation{
\wpauli
}

}

\newcommand{\michaelbarnes}{
\author{M.\ Barnes}
\affiliation{
\rudolphpeierls
}
\affiliation{
\culham
}

}

\newcommand{\felixparra}{
\author{F.\ I.\ Parra}
\affiliation{
\rudolphpeierls
}

}

\newcommand{\colinroach}{
\author{C. M. Roach}
\affiliation{
\culham
}

}

\newcommand{\stevecowley}{
\author{S. C. Cowley}
\affiliation{
\culham
}
\affiliation{
\blackett
}

}

\newcommand{\alexschekochihin}{
\author{A.\ A.\ Schekochihin}
\affiliation{
\rudolphpeierls
}

}

\newcommand{\myfig}[3][3.0in]{
\begin{figure}[htpd]
\includegraphics[width=#1]{#2}%
\caption{#3\label{#2}}%
\end{figure}
}

\newcommand{\figref}[1]{Fig. \ref{#1}}

\newcommand{\prt}{\mathrm{Pr}_t}

\begin{document}

\title{Transport Bifurcation Induced by Sheared Toroidal Flow in Tokamak Plasmas} 

\edmundhighcock
\michaelbarnes
\felixparra
\alexschekochihin
\colinroach
\stevecowley

\date{\today}

\begin{abstract}
First-principles numerical simulations
are used to describe a transport bifurcation in a differentially rotating tokamak plasma. 
Such a bifurcation is more probable in a region of zero magnetic shear 
than one of finite magnetic shear  
because in the former case
the component of the sheared toroidal flow that is perpendicular to the magnetic field 
has the strongest suppressing effect on the turbulence.
In the zero-magnetic-shear regime,
there are no growing linear eigenmodes at any finite value of flow shear. 
However, subcritical turbulence can be sustained, owing to the existence of modes, driven by the ion temperature gradient and the parallel velocity gradient, which grow transiently. 
Nonetheless, in a parameter space containing a wide range of temperature gradients and velocity shears, 
there is a sizeable window where all turbulence is suppressed.
Combined with the relatively low transport of momentum by collisional (neoclassical) mechanisms, this produces the conditions for a bifurcation from low to high temperature and velocity gradients.
A parametric model is constructed
which accurately describes the combined effect of
the temperature gradient
and the flow gradient 
over a wide range of their values.
Using this parametric model,
it is shown that in this reduced-transport state, heat is transported almost neoclassically, while momentum transport is dominated by subcritical PVG turbulence.
It is further shown that for any given input of torque,
there is an optimum input of heat which maximises the temperature gradient.
The parametric model describes both the behaviour of the subcritical turbulence
(which cannot be modelled by the quasi-linear methods used in current transport codes)
and the complicated effect of the flow shear on the transport stiffness.
It may prove useful for transport modelling of tokamaks with sheared flows.
\end{abstract}


\maketitle 

\section{Introduction}

A major obstacle to the development
of a working magnetic confinement fusion device
is the transport of heat
by pressure-gradient driven turbulence \cite{wolf2003internal}. 
Experimental results indicate that a shear
in the equilibrium plasma flow can reduce,
or even eliminate, this turbulence
\cite{burrell1997effects, wolf2003internal, connor2004itbreview}. 
Several numerical studies 
(gyrofluid \cite{waltz1994toroidal}, quasilinear \cite{waltz1997gyro},
and gyrokinetic \cite{dimits2001parameter, kinsey2005flowshear, camenen2009impact, roach2009gss, barnes2011turbulent, highcock2010transport})
have reproduced this effect and established that it results from the shear in the flow velocity component perpendicular to the equilibrium magnetic field. 
It has also been noted that a shear
in the flow velocity component parallel to the field,
which leads to a linear instability
\cite{catto1973parallel, newton2010understanding, schekochihin2010subcritical},
can have the opposite effect and increase the turbulence amplitudes
\cite{kinsey2005flowshear, roach2009gss, barnes2011turbulent, highcock2010transport}. 
Some numerical studies have concluded  that,
in the regimes considered,
when the effect of parallel flow shear is included
the turbulence is never fully suppressed
\cite{dimits2001parameter, kinsey2005flowshear}. 
This contrasts with experimental results
where almost neoclassical,
i.e., collisional \cite{rosenbluth1972plasma},
levels of heat transport have been observed
\cite{burrell1997effects, connor2004itbreview}. 

The ratio between the destabilising parallel
and stabilising perpendicular components of the flow shear
is set by the angle of the flow with respect to the magnetic field.
In general,
friction between trapped and passing particles 
and magnetic pumping keep the equilibrium flow nearly toroidal, 
with a large component of destabilising parallel flow shear
\cite{catto1987ion, hintonwong1985nit, flowtome1}.
Nonetheless, two recent papers 
\cite{barnes2011turbulent, highcock2010transport} 
which investigated values of the flow shear
several times larger than the linear growth rate 
of the ion temperature gradient (ITG) instability 
(one of the principal drivers of turbulence in fusion devices) 
found that for relatively low temperature gradients 
there is a range of flow shear values where  
\emph{a purely toroidal flow} can completely quench the turbulence.

The question remains how large shears in the equilibrium toroidal flow can be achieved, given finite available sources of angular momentum. 
In Refs. \cite{barnes2011turbulent, highcock2010transport}, 
the turbulent flux of angular momentum  was calculated and found to be large at moderate flow shears; 
if it is too large, increasing the input of angular momentum will not generate strong enough shear in the flow. 
Ref. \cite{barnes2011turbulent} considered the standard Cyclone Base Case \cite{dimits:969}, with the normalised inverse magnetic gradient scale length $\hat{s}=0.8$ and a range of values of the temperature gradient and the velocity shear. 
In this paper, which expands the results reported in \cite{highcock2010transport}, 
we consider a similar regime to that described in \cite{barnes2011turbulent},
with the difference that here the magnetic shear $\hat{s}=0$.  
This choice is motivated by the experimental studies
which have found that internal transport barriers 
(local regions of very steep flow and temperature gradients)
tend to form in regions of zero or low magnetic shear \cite{connor2004itbreview, vries2009internal}, 
and by previous numerical results which show a lower ion thermal diffusivity
at zero magnetic shear \cite{casson2009anomalous}.
The results reported below show that
the range of flow velocity gradients and ion temperature gradients
where the turbulence is suppressed
is much larger than in the case of finite magnetic shear
and, therefore, that a transition in both the flow and temperature gradients
from low to high values is more readily achieved
(such a transition is in principle also possible with \(\hat{s}=0.8\),
but in a much smaller region of parameter space \cite{parra2011momentum}).  
A positive feedback
between the increase in the flow gradient
and the suppression of the turbulence
provides the mechanism for a jump from low to high flow gradients,
with a simultaneous jump in the temperature gradient. 
We show that this transition results in a state
where the transport of heat is nearly neoclassical,
whilst the transport of momentum remains largely turbulent. 

In a fusion device, the quantities that can be controlled are the input of heat and toroidal angular momentum, which in steady state are proportional to their outgoing fluxes.
By varying these quantities, it is possible to vary the equilibrium gradients.
In contrast, in the local gyrokinetic simulations used in this study
the input parameters are the gradients, and the fluxes are calculated from the output. 
In order to demonstrate the existence of
a bifurcation in the gradients
at fixed values of the input fluxes,
we use local gyrokinetic simulations
to map out the dependence of the turbulent fluxes
on the temperature and flow gradients,
and then we add the neoclassical fluxes
and invert this numerically determined dependence
to find the gradients as functions of the total fluxes. 
We first do this by straightforward interpolation in the parameter space, then propose a simple parameterisation of the fluxes that fits the data and can be used to predict in which parameter regimes bifurcations can occur.

The rest of this paper is organised as follows.
In Section \ref{modelsec}, we present the gyrokinetic system of equations used in our simulations and describe the numerical setup.
In Section \ref{fluxessec}, we report a numerical scan in two parameters: the flow shear and the ion temperature gradient, calculating the turbulent heat and momentum fluxes over wide intervals of these parameters. 
In Section \ref{subcriticalsec}, we describe the subcritical turbulence, driven by the ion temperature gradient (ITG) and the parallel velocity gradient (PVG), that is responsible for the turbulent fluxes.
In Section \ref{transitionsec}, we interpolate our results and determine how the temperature and flow gradients depend on the heat and momentum fluxes: a transport bifurcation is obtained as a result.  
In Section \ref{qbehaviour}, we construct a parametric model of the transport. 
This allows us, in Section \ref{furtherobservationssec}, to study the effect of the transport bifurcation  on the temperature gradient and investigate the range of heat and momentum flux values for which we expect transitions to reduced transport to occur.
In Section \ref{conclusions} we summarise and discuss the results.

\section{Equations and Numerics}

\label{modelsec}

\subsection{Gyrokinetics with Velocity Shear}

We use the gyrokinetic approximation \cite{frieman1982nge}
to calculate the way in which the equilibrium gradients of the temperature and flow 
affect the transport of heat and momentum by turbulent fluctuations.
For a detailed description of high-flow gyrokinetics, 
the reader is referred to \cite{flowtome1, sugama1998neg, artun1992gai} and references therein. 
Here a brief summary is presented.

We consider a toroidal plasma with an axisymmetric equilibrium magnetic field $\vct{B}$. 
The field lines lie on closed flux surfaces,
which can be labelled by the poloidal magnetic flux $\psi$
contained within each surface. 
The magnetic field can be written as
$\vct{B} = \nabla \phi \times \nabla \psi + \btor R \nabla \phi$,
where \(\phi\) is the toroidal angle,  
$R$ is the radial coordinate measured from the central axis of the torus (the major radius),
and $\btor$ is the toroidal magnetic field. 
Owing to the fast motion of the particles along the magnetic field lines,
equilibrium quantities such as the density \footnote{
For the density, this is only true if the toroidal flow is subsonic,
which is what we assume below, see Eq. \ref{flowmagordering}.} 
and temperature are constant on each flux surface. 
In the current study, which considers the Cyclone Base Case regime \cite{dimits:969}, 
the magnetic flux surfaces are concentric and circular in cross-section, 
so that $\nabla \psi = \bpol R \nabla r$, 
where $r$ is the minor radius of the torus and $\bpol$ the poloidal magnetic field. 
Thus, $r$ may be used as the flux label, and the magnetic field may be written as $\vct{B} = \bpol R \nabla \phi \times \nabla r + \btor R \nabla \phi$.

We allow an equilibrium plasma flow $\vct{u}$, 
of the same order as the ion thermal velocity 
\begin{equation}
	v_{thi} = \sqrt{\frac{2T_i}{m_i}}, 
	\label{vthidef}
\end{equation}
where $T_i$ is the temperature of the ions and $m_i$ is their mass \footnote{
The definition of the thermal velocity 
given in \eqref{vthidef} is chosen in accordance with the analytical works
\cite{flowtome1, hintonwong1985nit} cited in this paper;
maintaining correspondence with \cite{barnes2011turbulent, highcock2010transport},
where the ion thermal velocity was defined as $\sqrt{T_i/m_i}$,
results in various factors of $\sqrt{2}$ in the normalisations 
of the output from simulations.}.
It can be shown that such a flow is toroidal, 
as any poloidal component will be quickly damped \cite{catto1987ion, hintonwong1985nit}. 
The flow can then be expressed as \(\vct{u}= \omega R^2 \nabla \phi\), 
where \(\omega\) is the angular velocity of the flow (which must be constant on a flux surface).

The state of each species $s$ of charged particles can be described by its distribution function, $f_s$,
whose evolution is given by the Vlasov-Landau equation:

\begin{equation}
	\pdt{f_s} + \vct{v}\cdot\nabla f_s + \ddt{\vct{v}}\cdot\pd{f_s}{\vct{v}} = C \lsq f_s \rsq, 
	\label{<++>}
\end{equation}
where $\vct{v}$ is the velocity coordinate and $C$ the collision operator. 
In order to describe the turbulent fluctuations which give rise to the transport, the Vlasov-Landau equation can be expanded by splitting $f_s$ into an equilibrium part $F_s$ and a perturbed part $\delta f_s$.
The latter is further split into averaged and fluctuating parts; so

\begin{equation}
	f_s = F_s + \left< \delta f_s \right> + \delta \tilde{f_s} 
	\label{deltafsdef}
\end{equation}
where the angle brackets denote a spatial and temporal average over fluctuations. 
It is assumed that the perturbations are small and that the gyrokinetic ordering \cite{frieman1982nge, flowtome1} holds, so that: 

\begin{eqnarray}
	\frac{\delta f_s}{F_s} \sim \frac{k\pa}{k\pp} \sim \frac{\gamma}{\Oi} \sim  \frac{\rho_i}{L} \equiv \epsilon \ll 1, 
	\label{orderings}
\end{eqnarray}
where
\(k\pa\) and \(k\pp\) are the typical parallel and perpendicular wavenumbers of the fluctuations,
\(\gamma\) is the growth rate of the fluctuations,
\(\Oi\) and $\rho_i$ are the gyrofrequency and Larmor radius of the ions, respectively,
and \(L\) is scale length of the variation of $F_s$.
It is then possible to show \footnote{
With the additional assumption
that the like-particle collision frequency $\nu_{ss}$
satisfies $\epsilon^2 \gamma \ll \nu_{ss} \lesssim \gamma$.}
that to lowest order,
$F_s$ is a local Maxwellian in the frame of the equilibrium flow $\vct{u}$,
and that, to first order in $\epsilon$,

\begin{equation}
	\delta \tilde{f_s}  =  - \frac{Z_s e\varphi}{T_s} F_s + h_s \lp t, \vct{R}_s, \mu_s, \varepsilon_s\rp
	\label{hsdef}
\end{equation}
where \(Z_s e\) is the charge of species $s$ and 
\(\varphi\) is the perturbed electrostatic potential. 
The non-Boltzmann part of the distribution function, $h_s$, is the gyrocentre distribution,
independent of the gyrophase and a function of time $t$, 
the gyrocentre position $\vct{R}_s$,
the magnetic moment $\mu_s$ 
and the energy $\varepsilon_s$ of the particle \cite{flowtome1}. 
The gyrocentre position $\vct{R}_s=\vct{r} - \uv{b}\times \vct{v} / \Omega_s $, where 
$\vct{r}$ is the position coordinate,
$\Omega_s$ is the gyrofrequency of species $s$,
the unit vector $\uv{b} = \vct{B}/B$ is the direction of the equilibrium magnetic field,
and $B$ is its magnitude.

The turbulence can be affected both by the gradient of the flow and by its magnitude, 
the latter through the Coriolis and centrifugal drifts. 
Here we neglect the effects of the magnitude of the flow
(the consequences of this simplification are discussed briefly in Section \ref{conclusions}) 
but allow the gradient of the flow to be of order the growth rate of the fluctuations,
$R d\omega / dr \sim \gamma$,
as is necessary for the flow gradient to affect the turbulence non-negligibly \cite{waltz1994toroidal}\footnote{
This can be systematised
by assuming
that the magnitude of the flow is small but the gradient of the flow is large 
in a Mach-number ordering subsidiary to \eqref{orderings}, namely 
\begin{equation}
	\frac{R\omega}{v_{thi}} \sim M \ll 1,\; 
	\frac{R}{\omega} \dbd{\omega}{r} \sim \oov{M} \gg 1
	\label{flowmagordering}
\end{equation}
where $\epsilon \ll M \ll 1$.
}.

Under these combined assumptions, it can be shown
that $h_s$ evolves according to the following gyrokinetic equation:

\begin{multline}
	\ddt{ h_s} + \lp v\pa \uv{b} + \vct{V}_D + \frac{c}{B}\uv{b}\times \nabla\gyroR{\varphi} \rp \cdot \nabla h_s =\\
	\frac{Z_s e}{T_s} \lsq  \ddt{ \gyroR{\varphi}}  - \frac{\btor v\pa R}{B \Omega_s}\dbd{\omega}{r} \lp \uv{b} \times \nabla 
\gyroR{\varphi}\rp \cdot \nabla r \rsq F_{s} \\
- \frac{cF_s}{B}  \lp \uv{b} \times \nabla \gyroR{\varphi} \rp \cdot \nabla r \lsq \oov{n_s} \dbd{n_s}{r} + \oov{T_s}\dbd{T_s}{r} \lp  \frac{\varepsilon_s}{T_s} - \frac{3}{2} \rp \rsq \\  + \gyroR{C\lsq h_s \rsq}, 
	\label{gyrokineticeq1}
\end{multline}
where $n_s$ and $T_s$ are the equilibrium density and temperature of species $s$, respectively,
$\gyroR{}$ is a gyroaverage at constant $\vct{R}_s$,
$d/dt = \partial / \partial t + \vct{u}\cdot \nabla$, 
\(\vct{V}_D\) is the magnetic drift velocity, 
$v_{\pa} = \sigma \sqrt{\lp \varepsilon_s - \mu_s B \rp/m_s}$ is the parallel (peculiar) velocity, 
$\sigma$ is the sign of $v_{\pa}$, 
$m_s$ is the mass of species $s$,
and $c$ is the speed of light.
We have further simplified the problem by assuming purely electrostatic perturbations ($\delta \vct{B}=0$),
so the system is closed by the quasineutrality condition:
\begin{equation}
	\sum_s \frac{  Z_s^2 e \varphi}{T_s}n_s = \sum_s Z_s \int d^3  \vct{v} \gyror{h_s},
	\label{quasin}
\end{equation}
where $\gyror{}$ is a gyroaverage at constant $\vct{r}$.
The electrons are treated through a modified Boltzmann response \cite{hammett1993developments}:

\begin{equation}
	\int d^3 \vct{v} \gyror{h_e} = \frac{e \lp \varphi - \overline{\varphi}\rp}{T_e} n_e,
	\label{boltzmanne}
\end{equation}
where the overbar denotes averaging over the flux surface.
Thus, Eq. \eqref{gyrokineticeq1} is only solved for ions: $s=i$.

Knowing \(h_i\), 
we can calculate the turbulent heat and angular momentum fluxes
associated with a given minor radius and given values of the gradients,

\begin{equation}
	Q_t =  \overline{\left< \int d^3 \vct{v} \oov{2} m_i v^2 \frac{c}{B}  \gyror{h_i}\lp \uv{b}\times \nabla \varphi \rp \cdot \nabla r \right>},
	\label{heatfluxdef}
\end{equation}

\begin{equation}
	\Pi_t = \overline{\left< \int d^3 \vct{v}  m_i \gyror{h_i\vct{v}} \cdot \nabla{\phi}  R^2\frac{c  }{B} \lp  \uv{b}\times \nabla \varphi \rp \cdot \nabla r \right>}.
	\label{momfluxdef}
\end{equation}

\subsection{Numerical Setup}

The nonlinear gyrokinetic equation \eqref{gyrokineticeq1} is solved  using the gyrokinetic code {\tt GS2} \cite{gs2ref, gyrokineticswebsite}. 
Its solution depends on (among other quantities) the equilibrium gradients:
$dn_i/dr$, $dT_i/dr$ and $(\btor/B)d\omega/dr$ 
(the latter of which is the gradient the of parallel component of the background flow).
To simplify the treatment of these gradients,
we consider a small region of the plasma known as a flux tube \cite{beer:2687}, 
which encompasses a set of magnetic field lines,  
extending to include several turbulence decorrelation lengths
in each direction.
This allows equilibrium quantities to be expanded 
around a reference flux surface near the centre of the domain, labelled by  $r_0$. 
Thus we may write:
\begin{align}
	n_i \lp r \rp = n_i \lp r_0 \rp + \lp r - r_0 \rp \left. \dbd{n_i}{r} \right|_{r=r_0} \nonumber\\ 
	\equiv n_{i0} \lsq 1 - \lp r - r_0 \rp \frac{1}{L_n}\rsq, \label{localngrad} \\
	T_i \lp r \rp = T_i \lp r_0 \rp + \lp r - r_0 \rp \left. \dbd{T_i}{r} \right|_{r=r_0} \nonumber\\
	\equiv T_{i0} \lsq 1 - \lp r - r_0 \rp \frac{1}{L_T}\rsq, \label{localTgrad}\\
	\omega \lp r \rp = \omega \lp r_0 \rp + \lp r - r_0 \rp \left. \dbd{\omega}{r} \right|_{r=r_0} \nonumber \\
	\equiv \omega_0  + \lp r - r_0 \rp \frac{q_0}{r_0}\fsr.
	\label{localomegagrad}
\end{align}
where $L_n$ is the equilibrium density scale length, $L_T$ the equilibrium temperature scale length.
The perpendicular velocity shear and the magnetic safety factor are:
\begin{equation}
	\fsr = \frac{r_0}{q_0}\left.\dbd{\omega}{r}\right|_{r_0},\qquad q_0 = \frac{r_0\btor}{R_0\bpol}
	\label{fsrdef}
\end{equation}
respectively, where $R_0$ is the major radius of the magnetic axis.
For a given equilibrium magnetic field, the solution of a flux tube simulation
and, therefore, the values of $Q_t$ and $\Pi_t$, is a function of the 
three parameters $L_n$, $L_T$ and $\fsr$.
Defining the local radial coordinate as 

\begin{equation}
	x =  r - r_0 , 
	\label{xdef}
\end{equation}
and the second perpendicular coordinate $y$ as

\begin{equation}
	y =\frac{\bpol R_0}{\btor}\lp \phi - q_0 \theta \rp ,
	\label{ydef}
\end{equation}
where $\theta$ is the poloidal angle\footnote{It should be noted 
that because the sign of $B_{\theta}=\vct{B}\cdot\nabla\theta$ 
in this paper is opposite to 
the sign of the poloidal field in {\tt GS2},
the coordinate $y$, defined in \eqref{ydef},
also has the opposite sign to the {\tt GS2} coordinate of the same name.},
then transforming to a frame rotating with angular frequency $\omega_0$
and keeping terms to the correct order in the gyrokinetic expansion, we find for any perturbed quantity $\xi$:

\begin{eqnarray}
	\ddt{\xi} = \pdt{\xi} + \vct{u}\cdot \nabla \xi = \pdt{\xi}+ x \fsr \pd{\xi}{y}.
	\label{fsrequiv}
\end{eqnarray}
The gyrokinetic equation \eqref{gyrokineticeq1} may now be written

\begin{multline}
	\pdt{ h_i} + x\fsr\pd{h_i}{y} + \lp v\pa \uv{b} + \vct{V}_D + \frac{c}{B}\uv{b}\times \nabla\gyroR{\varphi} \rp \cdot \nabla h_i =\\
	\frac{Z_i e}{T_{i0}} \lsq  \pdt{ \gyroR{\varphi}} + x\fsr\pd{\gyroR{\varphi}}{y} 
	+ \right.  \left. \frac{\btor v\pa R}{B \Omega_i}\frac{q_0 \fsr}{r_0} \pd{\gyroR{\varphi}}{y}
	\rsq F_{i} \\
	- \frac{cF_i}{B}  \pd{ \gyroR{\varphi}}{y} \lsq \oov{L_n} + \oov{L_T} \lp \frac{\varepsilon_i}{T_{i0}} - \frac{3}{2} \rp \rsq \\  + \gyroR{C\lsq h_i \rsq}.
	\label{gyrokineticeq}
\end{multline}

In this investigation, we 
vary
the perpendicular flow shear $\fsr$,
and $\kappa = R_0/L_T$.
Other parameters
are kept fixed:
\begin{equation}
	\frac{R_0}{L_n}=2.2,\; q_0 = 1.4,\; \frac{r_0}{R_0}=0.18.
	\label{params}
\end{equation}
The density gradient, the magnetic safety factor and the aspect ratio
are chosen to conform to the Cyclone Base Case,
following \cite{dimits:969} and \cite{highcock2010transport},
while, as we explained above, the magnetic shear is set to 0.

Note that the value of $q_0$ effectively controls the strength of the PVG drive
(the term proportional to $q_0 \fsr$ on the right hand side of \eqref{gyrokineticeq})
for a given velocity shear $\fsr$.
The relatively low value of $q_0$ in the Cyclone Base Case
reduces the destabilising effect of the PVG
--- compared for example to the Waltz Standard Case where $q_0=2.0$ \cite{kinsey2005flowshear}.

Collisions are included by means of a model collision operator,
which includes the scattering of particles
in both pitch angle and energy \cite{numerics, abel2008linearized}. 
The numerical ion-ion collision frequency  $ \nu_{ii}^N = 0.01 v_{thi}/R_0$.

The effects of the perpendicular flow shear are included in the code as follows \cite{gs2flowshear}. 
Expanding
$\xi = \sum_{k_x, k_y} \hat{\xi} (z, t) \exp \lsq i\lp k_y y + k_x x \rp \rsq$, 
where $z$ is the coordinate along the field line, 
the perpendicular flow shear terms in \eqref{gyrokineticeq}
can be eliminated
by adding time variation to the radial wavenumber:

\begin{equation}
	k_x(t) = k_{x0} - \fsr k_y t, 
	\label{radialshear}
\end{equation}
so that

\begin{eqnarray}
	\pdt{\xi}+ x \fsr \pd{\xi}{y} = \sum_{k_{x0}, k_y}  \pdt{\hat{\xi}}  e^{i\lsq k_x\lp t\rp x + k_y y\rsq}.
	\label{eliminateperpterm}
\end{eqnarray}

All fluxes will be reported in dimensionless units
by normalising them to the 
gyro-Bohm values 
\begin{equation}
	Q_{gB}=\frac{n_iT_i v_{thi} \rho_i^2}{2 \sqrt{2} R_0^2},\qquad  \Pi_{gB}=\frac{m_i n_i v_{thi}^2 \rho_i^2}{4 R_0}.
	\label{gpdefs}
\end{equation}
The flow shear will be everywhere normalised to $ v_{thi} / R_0$;
hence at this point we rescale $\fsr$ accordingly:

\begin{equation}
	\fsr \rightarrow \frac{v_{thi}}{R_0} \fsr.
	\label{fsrnorm}
\end{equation}

The resolution of the majority of simulations was \(64\times32\times14\times24\times8\times2\)
--- 
the number of gridpoints in the $k_{x0}$ and $k_y$ spectral directions,
in the $z$ spatial direction,
the average number of pitch angles (i.e. the number of values of  $\mu_i$ for a given $\varepsilon_i$, which varies with the parallel coordinate),
the number of gridpoints in energy space $\varepsilon_i$
and the two values of $\sigma$. 
This grid provided sufficient scale separation
in both spatial directions 
perpendicular to the magnetic field
for calculating the turbulent transport,
and sufficient resolution in velocity space
to calculate the velocity integrals in Maxwell's equations 
with the required accuracy. 
A discussion of the parallel resolution is given in Section \ref{subcriticalsec}.

\section{Turbulent Fluxes}
\label{fluxessec}
\subsection{Heat Flux}
Using simulations in the manner described in Section \ref{modelsec}, 
the heat and momentum fluxes were calculated 
for $\tprim$ values in the range $4 \leq \tprim \leq 13$, 
vs flow shear values in the range $0\leq\fsr\leq2\gstwofsrnorm$.

\figref{gexbzero}(b) shows the heat flux vs the ITG in the absence of flow shear. 
When \(\fsr=0\), the critical temperature gradient
above which there is ITG-driven turbulence is \(\tprim=4.4\). 
When $\tprim$ is increased above this threshold,
the heat flux increases rapidly up to more than a hundred times the gyro-Bohm value.

\figref{fluxes} shows the turbulent heat and momentum fluxes vs the flow shear for different ITG values. 
As the flow shear is increased from 0, the heat flux initially responds weakly, 
either increasing or decreasing slightly.
As the flow shear $\fsr$ approaches 1, the heat flux dips sharply for all values of the ITG. 
For moderate temperature gradients (\(\tprim \lesssim 11\)),
the turbulence is suppressed altogether and the heat flux drops to zero. 
The suppression of turbulence also happens at finite magnetic shears \cite{barnes2011turbulent}, 
but for a significantly narrower range of $\tprim$ \footnote{
This is partly owing to the fact that at finite magnetic shear growing linear eigenmodes exist for non-zero flow shear, whereas at zero magnetic shear there are no such eigenmodes except when the flow shear is also zero; this is discussed further in Section \ref{subcriticalsec}.
}.

For larger $\fsr$, the heat flux starts to rise again. 
This increase is due to the PVG
--- we have verified that this effect disappears
if the term proportional to $q_0\fsr$
on the right hand side of eq \eqref{gyrokineticeq}
is artificially set to 0.
This revival of the turbulent transport at large shears
was also observed in \cite{barnes2011turbulent}, 
but was absent from the quasilinear study conducted in \cite{waltz1997gyro}, 
which showed the turbulence being completely suppressed above a sufficiently large toroidal shear.

\myfig{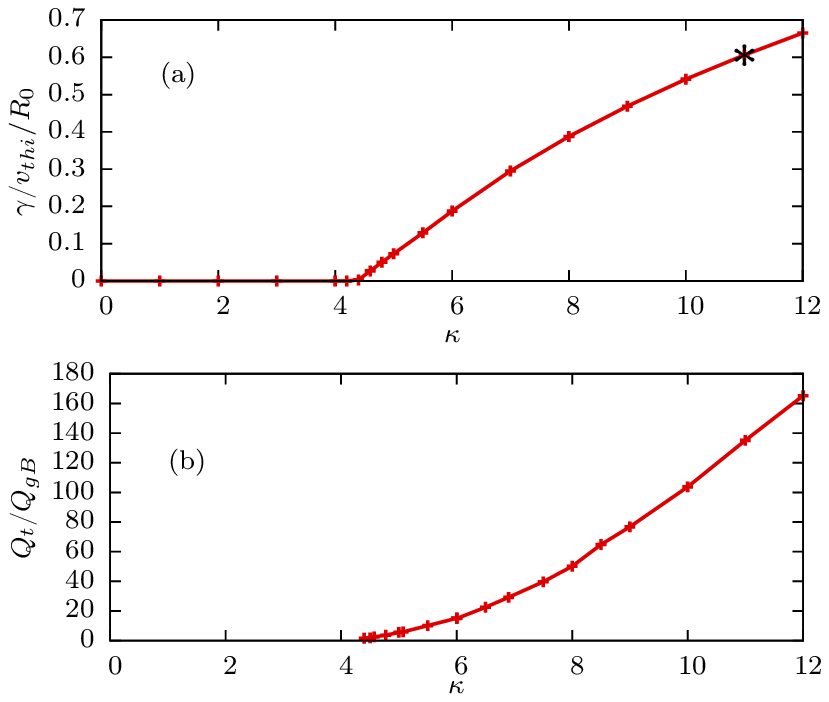}{ITG instability and turbulence at zero flow shear: (a) linear growth rate; (b) turbulent heat flux vs the normalised temperature gradient $\tprim=R_0/L_T$. The case of $\kappa=11$, further explored in \figref{transientgrowth}, is marked by *.}
\myfig{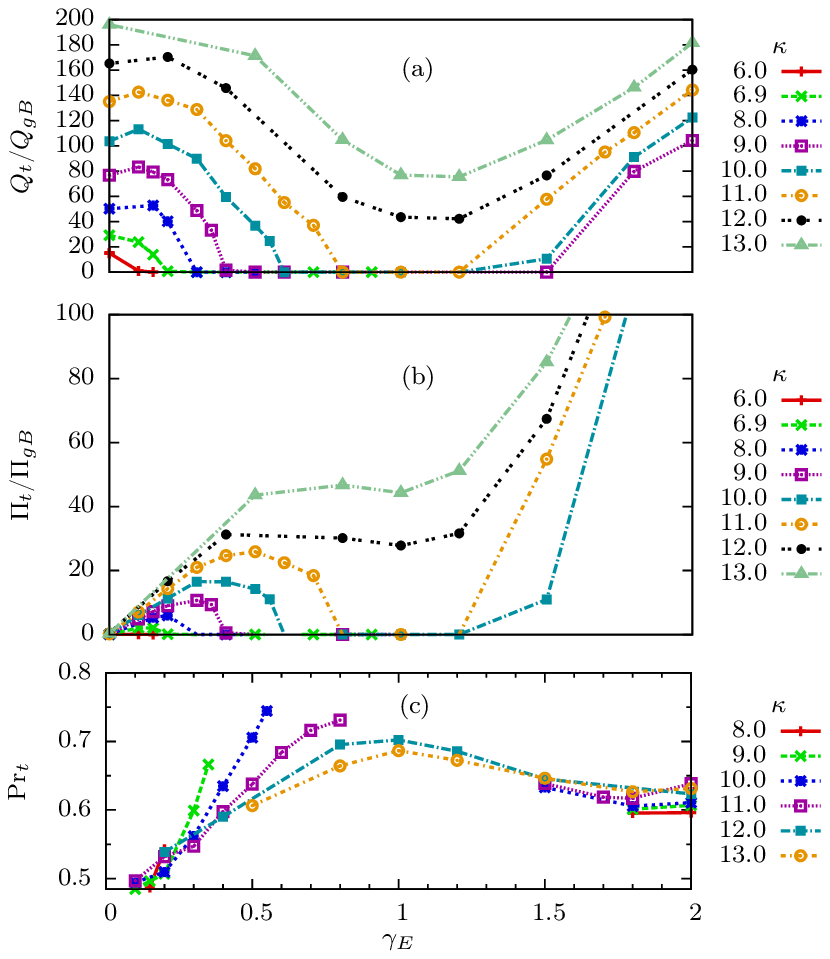}{Turbulent heat (a) and toroidal angular momentum (b) fluxes (normalised to gyro-Bohm values) as functions of flow shear for different values of the ion temperature gradient $\tprim$; (c) turbulent Prandtl number vs flow shear (for cases where the heat flux is non-zero).}

\subsection{Momentum Flux}

The momentum flux is zero when \(\fsr=0\) 
(as might be expected in an up-down symmetric case, 
\cite{parra2011up, camenen2009intrinsic}). 
As $\fsr$ increases, so does the momentum flux ---
a result of the turbulent viscosity.
However, when the flow shear reaches values at which it starts to suppress turbulence significantly, 
the trend is reversed and a remarkable situation arises 
in which increasing flow shear reduces the transport of momentum.
This behaviour persists until $\fsr$ reaches even larger values, 
and the turbulence is reignited by the PVG drive.
The positive correlation between the velocity shear and the momentum flux is then reestablished. 
It is the existence of the window of suppressed momentum transport 
at moderate $\fsr$ and $\tprim$ that will enable
the transport bifurcation analysed in Section \ref{transitionsec}.

\subsection{Turbulent Prandtl Number }
\label{prtsec}

There is a clear correlation between the heat and the momentum flux,
which is best quantified in terms of the turbulent Prandtl number

\begin{align}
	\mathrm{Pr}_t = \frac{\nu_t}{\chi_t} 
	= \frac{\Pi_t/\Pi_{gB}}{Q_t/Q_{gB}}\frac{\tprim}{\fsr} \frac{r_0}{\sqrt{2} R_0 q_0}
	\label{prtdef}
\end{align}
where the turbulent viscosity and the turbulent heat diffusivity are

\begin{align}
	&\nu_t = \frac{\Pi_t}{\Pi_{gB}} \frac{1}{\fsr} \frac{v_{thi} r_0  \rho_i^2}{4 q_0 R_0^2},\label{nutdef}\\
	&\chi_t = \frac{Q_t}{Q_{gB}}\oov{\tprim}\frac{v_{thi} \rho_i^2}{2\sqrt{2}R_0},\label{chitdef}
\end{align}
respectively.

The Prandtl number obtained from our simulations is plotted in Fig. \ref{fluxes}(c). 
There is a similar basic trend for all values of $\kappa$:
$\prt$ rises from approximately 0.5  when $\fsr\simeq0.1\gstwofsrnorm$ \footnote{
The low value of $\prt$ for small $\fsr$
has been observed before
and is sometimes referred to as a shear pinch 
\cite{dominguez1993anomalous,waltz2007gyrokinetic}.
It occurs because, at low $\fsr$,
the perpendicular flow shear can 
give rise to a contribution to the viscous stress 
that has the opposite sign to that of ITG-driven turbulence with zero flow shear,
reducing the overall diffusive transport.},
peaks at $\fsr\simeq1\gstwofsrnorm$ and then drops to approximately 0.6 when $\fsr\simeq2\gstwofsrnorm$. 
For those intermediate values of $\fsr$
where the turbulent fluxes are reduced or tend to 0,
$\prt$ rises sharply,
reaching just above 0.7  for low values of $\tprim$.
The location of this sharp rise varies with temperature gradient
similarly to the location where the turbulence is suppressed.

Although the Prandtl number does vary with both $\fsr$ and $\tprim$,
this dependence is relatively weak
compared to the individual dependence of $\nu_t$ and $\chi_t$ on these parameters.
Thus,  approximating $\prt\sim0.55$ everywhere 
keeps it within approximately 25\% of the true value for the entire range of $\fsr$.
This will prove convenient in constructing a model of turbulent transport presented in Section \ref{qbehaviour}.

\section{Subcritical Turbulence}
\label{subcriticalsec}

Before studying the implications
of the results presented above, 
we first discuss the nature of the fluctuations
that give rise to the behaviour of
the turbulent fluxes discussed in Section \ref{fluxessec}.

\myfig{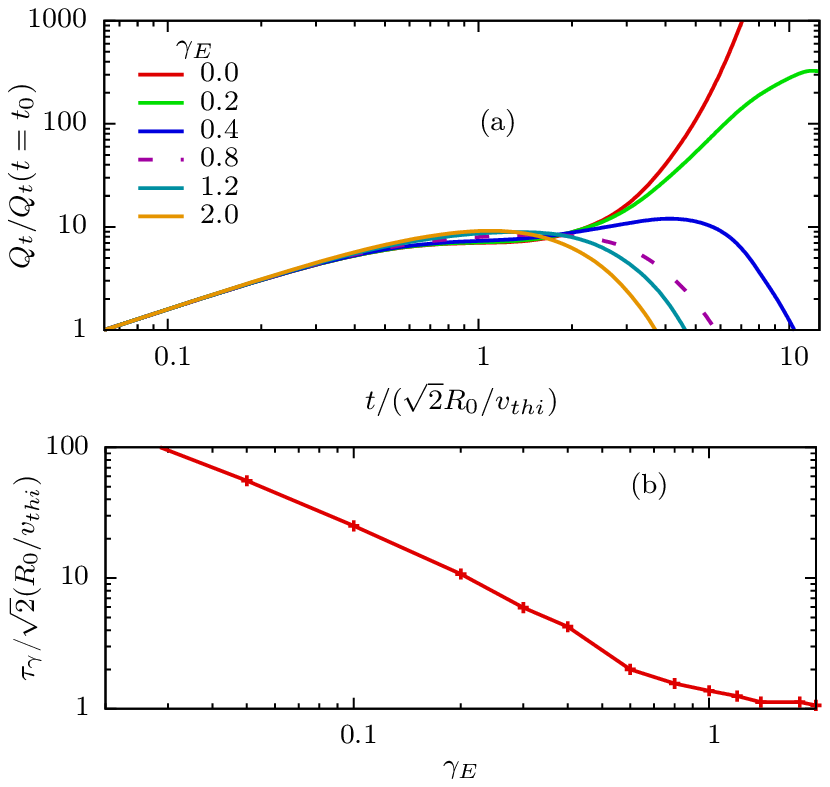}{Transiently growing linear modes at \(\tprim=11\).
(a) The heat flux as a function of time, normalised to its value at $t_0$,
where $t_0=0.06\sqrt{2}R_0/v_{thi}$,
so chosen to skip the short initial transient associated with a particular choice of 
initial condition.
All modes grow transiently for \(\fsr>0\). (b) Duration of transient growth \(\tau_{\gamma}\) from $t=t_0$ to the peak value of $Q_t$ vs flow shear.}
\myfig{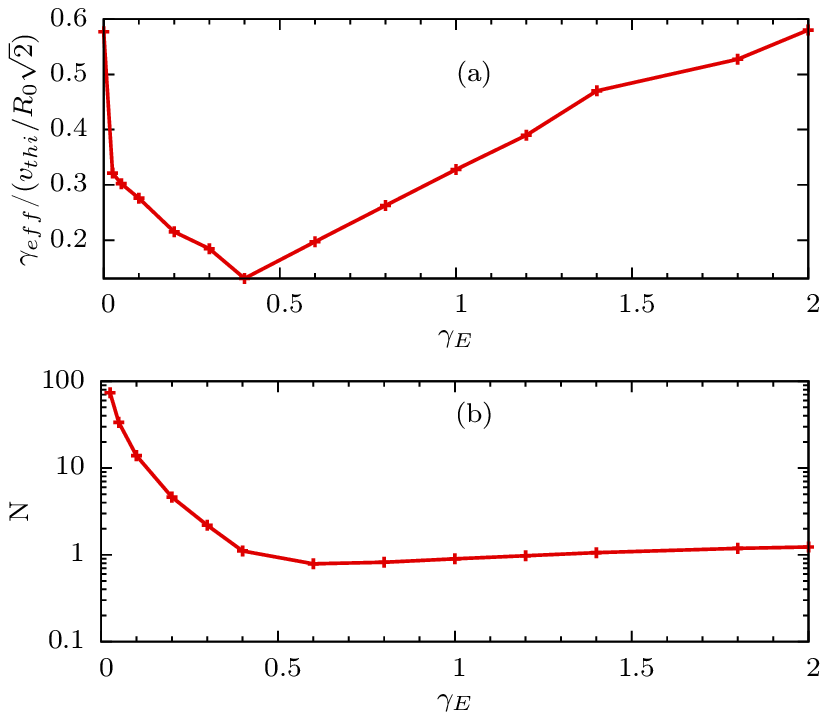}{Two measures of the strength of the transiently growing linear modes at $\kappa=11$: (a) the effective growth rate of the mode and (b) the number of exponentiations of the heat flux during the growing phase, both vs. the flow shear. 
The apparent discontinuity at $\fsr=0$ in (a) is a result of taking the average growth rate of the heat flux 
over $0 < t < \tau_{\gamma}$ rather than the initial or peak growth rate; 
for all $\fsr>0$ the average will include a period where the growth rate tends to zero.
}

With \(\fsr=0\), the ITG instability exhibits robust linear growth, 
with a threshold of \(\kappa=4.4\) (Fig. \ref{gexbzero}(a)).
However, for any finite value of the flow shear 
there are no growing linear eigenmodes in the system:
all modes grow only transiently before decaying (Fig \ref{transientgrowth}(a)). 
While formally this means that $\fsr\rightarrow0$ is a singular limit,
there is is in fact no physical discontinuity in behavior:
as $\fsr \rightarrow0$, the duration of the transient growth,
$\tau_{\gamma}$, tends to infinity (roughly as $1/\fsr$; see Fig. \ref{transientgrowth}(b)). 
There is therefore a continuous transition at $\fsr=0$
from a transient mode that grows for 
an infinitely large time, to a growing eigenmode.
The lack of growing eigenmodes when $\fsr>0$ stems from the wavenumber drift \eqref{radialshear}: 
as time tends to infinity at finite flow shear and zero magnetic shear, 
the radial wavenumbers increase inexorably through the shearing of the modes, 
until they are extinguished by finite-Larmor-radius effects. 

This situation differs somewhat from the case with finite magnetic shear $\hat{s}$ 
\cite{barnes2011turbulent, artun1993integral}, 
where there are linearly growing modes for a finite range of \(\fsr\). 
At finite magnetic shear,
growing eigenmodes are possible 
because while $\fsr$ leads to an effective dependence of the radial wavenumbers on time, 
$\hat{s}$ makes them dependent also on the position along the field line.
Therefore, for modes moving with a speed proportional to $\fsr/\hat{s}$, 
the magnetic shear cancels the effect of the velocity shear on $k_x\lp t\rp$ 
\cite{newton2010understanding, waltz1994toroidal}. 
When the magnetic shear becomes very small, or the flow shear very large,
the required mode velocity becomes much greater than the thermal speed of the ions 
and the cancellation is no longer possible 
because the mode cannot travel so fast \cite{newton2010understanding}.
As a consequence, growing linear eigenmodes only exist for non-zero magnetic shear
and only up to a finite value of the flow shear \cite{barnes2011turbulent}.

In a standard picture of ITG turbulence with $\fsr=0\gstwofsrnorm$ \cite{dimits:969},
the turbulence is driven by unstable linear modes.
With the exception of a narrow interval of temperature gradients
where self-generated zonal flows suppress the turbulence (the Dimits shift),
the presence and the amplitude of the turbulence
are largely determined by the presence and the growth rate of those unstable modes.
In the present case, by contrast, we have strong levels of turbulence sustained nonlinearly 
in a parameter regime where there are no linearly unstable modes, 
a phenomenon first discovered in tokamak turbulence 
in \cite{barnes2011turbulent} and \cite{highcock2010transport}, 
but well known in various hydrodynamical contexts \cite{trefethen1993hydrodynamic}. 
This phenomenon is known as \emph{subcritical turbulence}.

Subcritical turbulence differs from standard instability-driven turbulence in several important ways. 
Firstly, because there is no linear instability, 
turbulence will not grow from initial perturbations of arbitrarily small amplitude 
\cite{barnes2011turbulent}. 
Fluctuations must be initialised with sufficient amplitude (generally of the order of their amplitude in the saturated state) in order for turbulence to be sustained; 
thus, the absence of sustained turbulence in a numerical experiment may merely indicate that the initial amplitude is insufficient, 
not that the plasma is quiescent. 

The second difference concerns the scales at which
the turbulent energy resides.
In ITG-driven supercritical turbulence ($\fsr=0$),
these scales are those where the linear drive injects energy ---
this tends to correspond to $k_{\pa} q_0 R_0 \sim 1$
and $k_y \rho_i$ relatively narrowly concentrated
around values of a fraction of unity
(at least for low values of $q_0$ and moderate temperature gradients;
see \cite{barnes2011flucspec} and \figref{subcriticalspectrum}(a)).
In subcritical turbulence,
the preferred wavelengths appear to be
those at which the amplification of the transient modes is strongest.
In the case of turbulence where the PVG drive is dominant, 
Ref. \cite{schekochihin2010subcritical} has shown that the amplification 
is maximised along a line in Fourier space where 

\begin{equation}
	k_y \rho_i \simeq  \lp \frac{r_0}{R_0q_0} \rp^{1/3}\frac{k\pa R_0}{\sqrt{2}\fsr} \gstwofsrnorm.
	\label{maxamp}
\end{equation}
Fig. \ref{subcriticalspectrum}(b) shows the spectrum 
of strongly PVG-driven turbulence at high flow shear ($\fsr=2.2\gstwofsrnorm$).
We see that although the result \eqref{maxamp} 
is based on linear theory and
is derived for slab geometry,
it appears to describe the peak of the spectrum reasonably well. 
The spectrum extends to much higher 
parallel wavenumbers than in the standard ITG case; 
consequently, higher parallel resolution is necessary to resolve it\footnote{Studies of the effect of increasing the parallel resolution showed that at values of flow shear $\fsr\gtrsim 1.2$, the parallel resolution used for the parameter scan described in Section \ref{fluxessec} (14 gridpoints) was insufficient and led to errors in the critical temperature gradient (above which subcritical turbulence could be sustained) of order 5-10\%. 
We do not consider this error
significant enough to merit a repeat of the parameter scan
with substantially higher parallel resolution
(although in principle such an exercise would be useful)
.}.

Finally, faced with subcritical turbulence,
we are left without an intuitively obvious way of estimating its saturation level
and, consequently, the turbulent fluxes. 
It is the presence of linear eigenmodes with a defined growth rate (positive or negative)
which has enabled the quasilinear modelling of the heat flux 
in many situations
without resorting to full nonlinear simulations
(\cite{drummond1964nonlinear} was an early exposition; see \cite{krommes2002fundamental} for an overview and \cite{bourdelle2007new, waltz2009gyrokinetic} for recent work).
When the growth of all modes is transient, 
such models will not work in their current form.
The question arises as to which characteristics
of the linear transient growth
are relevant in the resulting nonlinear state.
The problem is as yet unsolved, but here we consider
two natural measures of subcritical driving.

Let us first
define the effective transient growth rate \cite{roach2009gss} 

\begin{equation}
	\gamma_{\mathrm{eff}}=\frac{1}{2 \tau_\gamma}\mathrm{ln} \frac{Q_t(t=\tau_\gamma)}{Q_t(t=t_0)} .
	\label{gammaeffdef}
\end{equation}
This effective growth rate is plotted in \figref{transientstrength}(a).
It initially decreases with increasing flow shear, 
but then starts to increase for $\fsr\gtrsim0.4\gstwofsrnorm$ 
as the PVG drive becomes significant. 
The effective growth rate seems to correlate
with the behaviour of the heat flux (Fig. \ref{fluxes}(a)):
it first decreases and then increases with increasing flow shear. 
However, the minimum in the effective growth rate
(for $\tprim=11$ as shown in Fig. \ref{transientstrength}(a))
occurs at \(\fsr\sim0.4\gstwofsrnorm\)
whereas the minimum in the heat flux for the same $\tprim$ occurs at $\fsr\sim1$.

Besides the rate of growth,
what is likely to be decisive in determining whether
the finite transient growth of perturbations
proves sufficient to maintain a steady level of turbulence is
the time that this growth lasts.
A natural diagnostic that accounts for both effects is 
the total transient amplification of the modes before they start to decay \cite{schekochihin2010subcritical, trefethen1993hydrodynamic, baggett1995mostly}:
\begin{equation}
	N=\tau_{\gamma}\gamma_{\mathrm{eff}}.
	\label{ndef}
\end{equation}
It drops precipitously as $\fsr$ is increased from 0, 
reaches a minimum at $\fsr=0.6$
(which does not coincide with $\fsr=1$, where $Q_t$ is minimal)
and then appears to increase very gently with $\fsr$
\footnote{This behaviour approximately agrees with the analytical result of \cite{schekochihin2010subcritical},
which predicts $N\rightarrow0.225q_0 R_0/r_0\sim1.75$ as $\fsr\rightarrow\infty$.}
---
in sharp contrast with the very rapid increase of
the nonlinear fluxes at high flow.

Thus, while the linear behaviour shows that the nonlinear fluxes 
are somewhat determined by the vigour of the underlying linear drive, 
a quantitative model of subcritical PVG turbulence 
that would allow the heat fluxes to be predicted
from the characteristics of transient linear growth
remains an unsolved problem.

\myfig{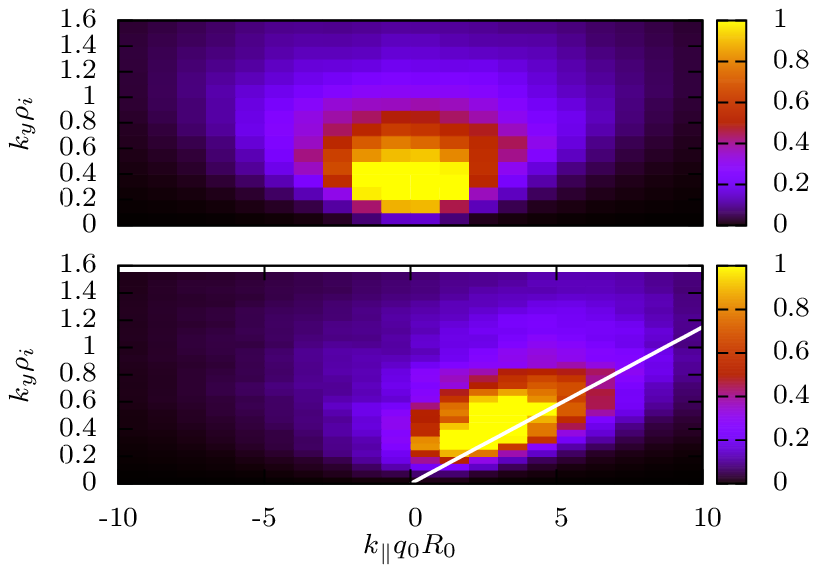}{The spectrum of turbulent fluctuations for normal and subcritical turbulence: \(k_y^2 \sum_{k_{x0}} \md{\hat{\varphi}_{k_{x0},k_y}}^2 \) (arbitrary units) vs \(k_y\) and \(k\pa\) for (a) \(\fsr=0.0,\,\tprim=12\) (ITG turbulence with no flow shear) and (b) \(\fsr=2.2,\,\tprim=12\) (subcritical, strong PVG-driven turbulence)\footnote{It should be noted that as the sign of $k_y$ is opposite in {\tt GS2} to the present work, these graphs were plotted via a parity transformation.}. Also shown in (b) is the line of maximum transient amplification \eqref{maxamp}, as calculated in \cite{schekochihin2010subcritical}.}

\section{Transport Bifurcation}
\label{transitionsec}
\subsection{Possibility of Bifurcation}
In Section \ref{fluxessec},
we demonstrated the existence of a wide interval in $\fsr$
in which a sheared toroidal equilibrium flow completely suppresses turbulent transport. 
If such a suppression could be achieved experimentally in a tokamak,
confinement of energy would be dramatically improved. 
Unfortunately, while it is possible to specify the flow shear in numerical simulations,
in experiment it can only be varied indirectly by applying torque,
and the effect of that torque is strongly dependent
on how quickly the angular momentum escapes from the plasma.
If only a limited amount of torque can be injected and the momentum flux is too large,
it could be impossible to achieve flow shears that are large enough to suppress the turbulence.  
Fig \ref{fluxes}(b), however, suggests an intriguing possibility. 
For all temperature gradients, the momentum flux at first increases,
reaches a maximum and then decreases
before increasing again at higher flow shears.
There is, therefore, a parameter window in which
increasing the flow shear decreases the transport of momentum. 
If this were to happen, the momentum would build up,
increasing the flow shear, which would further decrease the transport of momentum. 
Such a positive feedback could lead to a bifurcation and so it might seem that high values of flow shear could be reached without excessive input of momentum,
a possibility which was discussed in \cite{waltz1995advances} \footnote{
Note that although a similar mechanism was discussed in \cite{waltz1995advances}, 
that study considered flow shears
up to a maximum of approximately $0.08 c_s / a$, 
where $c_s=\sqrt{2T_e/m_i}$ is the sound speed 
and $a$ is the minor radius of the plasma (in this study $c_s=v_{thi}$ and $a=R_0$).
Thus they were considering a maximum in the momentum flux
which occurs at low flow shear and finite magnetic shear,
and which is discussed in more detail in \cite{barnes2011turbulent}.
As can be seen from Ref. \cite{barnes2011turbulent}, 
the corresponding jump in flow shear is much smaller,
and the range of $\tprim$ values 
at which such a maximum in the momentum flux exists is much narrower,
than in the present case (\figref{fluxes}(b)).
}. 
However, Fig \ref{fluxes}(b) shows the momentum flux \emph{at constant ion temperature gradient},
which would not, in fact, stay constant during this process. 
Indeed, as soon as flow shear began to suppress the turbulent transport of momentum, it would also suppress the turbulent transport of heat, causing an increase in the temperature gradient as well as in the flow gradient. 
This increase in the temperature gradient would restore the turbulence to its former levels.
Nevertheless, we will show in this section that when neoclassical transport is also taken into account, a bifurcation is possible.

\subsection{Inverting the Problem}
\label{invertingsec}
What can actually be controlled in a steady-state experimental situation
is the flow of heat and momentum through a particular surface.  
This means that we must switch from using \(\fsr\) and \(\tprim\) as independent parameters 
to using the total fluxes of heat and momentum, $Q$ and $\Pi$ respectively. 

Let us consider an experimental set-up
where the heat and the momentum are injected by beams of neutral particles.
We assume that the energy and momentum from those neutral beams
are deposited uniformly across the torus.
We further assume that the beams are tangential to the magnetic axis and
we ignore corrections of order the aspect ratio of the device.
Then:

\begin{align}
	&\frac{Q}{Q_{gB}}\simeq \oov{Q_{gB}} \frac{V_f N_b  m_i v_b^2/2}{4\pi^2 r_0 R_0}
	= V_f \frac{2\sqrt{2} R_0 P_{\mathrm{NBI}}}{4\pi^2 r_0 n_i T_i v_{thi} \rho_i^2},\,
	\label{qexp}\\
	&\frac{\Pi}{\Pi_{gB}} \simeq \oov{\Pi_{gB}}  \frac{V_f N_b m_i v_b}{4\pi^2  r_0},\, \\
	&\frac{\Pi/\Pi_{gB}}{Q/Q_{gB}} \simeq \frac{v_{thi}}{\sqrt{2}v_b} = \lp\frac{T_i}{2 E_{\mathrm{NBI}}}\rp^{1/2}, \,
	\label{pioqexp}
\end{align}
where $N_b$ is the number of neutral beam particles injected per unit time,
$v_b$ is the beam particle velocity,
$P_{\mathrm{NBI}}=N_b m_i v_b^2 / 2$ is the beam power,
$E_{\mathrm{NBI}}=m_i v_b^2 / 2$ is the beam particle energy
and $V_f$ is the volume fraction of the plasma enclosed by the flux surface.
Thus, the total heat flux $Q$ is determined by the beam power $P_{\mathrm{NBI}}$ \eqref{qexp},
whereas the ratio of the total momentum angular flux to the total heat flux,
\(\Pi/Q\), is determined by the beam particle energy $E_{\mathrm{NBI}}$ \eqref{pioqexp}. 
We want to know whether, by varying $Q$ and $\Pi/Q$,  it is possible to reach,
or to trigger a transition to, a high-flow-shear regime where the turbulent transport is suppressed. 
Thus, it is necessary to
convert the dependence of $Q$ and $\Pi$ on $\tprim$ and $\fsr$,
determined from local simulations, 
to a dependence of $\tprim$ and $\fsr$ on $Q$ and $\Pi/Q$.

\subsection{Interpolation}
\label{interpolationsec}
The gyrokinetic code {\tt GS2} gives the fluxes as a function of $\tprim$ and $\fsr$. 
Given the expense of the nonlinear simulations necessary to do this,
it is computationally challenging to use {\tt GS2}
as a root finder to invert the problem and find the gradients as a function of the fluxes
\cite{barnes2009trinity, candy2009tgyro}.
Instead we interpolate within the set of data points
described in Section \ref{fluxessec}
(as well as additional simulations in parameter regions of particular interest)
to obtain the fluxes as continuous functions of the gradients;
these functions can then be inverted numerically to give
$\tprim$ and $\fsr$ as functions of the fluxes.

Interpolation in a multidimensional parameter space
is a nontrivial operation.
A standard technique is to use radial basis functions \cite{buhmann2001radial},
which weigh each data point
by its distance in parameter space from the point of interest
(after the parameter space has been normalised
to ensure that variation occurs on the same scale in each coordinate).
There are many choices of the function, or kernel,
which is used to calculate the relative importance of each data point.
Here we choose a linear kernel
(equivalent to linear interpolation in the case of only two points) \cite{buhmann2001radial}. 
Using this, the values of the fluxes at a point
\( \vct{p} = (\fsr, \tprim) \) can be calculated as follows:

\begin{align}
	Q_t\lp \vct{p} \rp = \sum_{\mathrm{j}} w_\mathrm{j}^Q \md{\vct{p} - \vct{p}_{\mathrm{j}}},\\
	\Pi_t\lp \vct{p} \rp = \sum_{\mathrm{j}} w_\mathrm{j}^\Pi \md{\vct{p} - \vct{p}_\mathrm{j}},
	\label{interpolation}
\end{align}
where \(\vct{p}_\mathrm{j}\) are the input gradients for the nonlinear simulation labelled $\mathrm{j}$,
and the weights \(w_\mathrm{j}^Q\) and \(w_\mathrm{j}^\Pi\) are calculated
so as to satisfy for all $\mathrm{j}$:
\begin{align}
Q_t(\vct{p}_\mathrm{j}) = Q_{t\mathrm{j}}\;, \Pi_t(\vct{p}_\mathrm{j}) = \Pi_{t\mathrm{j}}, 
	\label{weightcalc}
\end{align}
where $Q_{t\mathrm{j}}$ and $\Pi_{t\mathrm{j}}$ are the values of the fluxes obtained from simulation $\mathrm{j}$.

\subsection{Neoclassical Transport}
\label{neoclassicaltransportsec}

If the turbulent transport is successfully suppressed,
neoclassical (collisional) transport becomes important. 
The total fluxes are the sum of the neoclassical and turbulent contributions: 

\begin{align}
	 Q = Q_t+Q_n,\;\Pi = \Pi_t + \Pi_n,
	 \label{totalqp}
\end{align}
where the neoclassical fluxes are calculated as
\begin{align}
	&\frac{Q_n}{Q_{gB}}=\chi_n\tprim\frac{2 \sqrt{2}R_0  }{v_{thi} \rho_i^2 },
	\\
	& \frac{\Pi_n}{\Pi_{gB}} =\nu_n\fsr \frac{4  R_0^2 q_0  }{r_0 v_{thi} \rho_i^2},
\end{align}
and the neoclassical thermal diffusivity and viscosity are
\begin{align}
&	\chi_n\simeq 0.66 \lp\frac{ R}{r_0}\rp^{3/2}q_0^2\rho_i^2 \nu_{ii},
	\label{chindef}
	\\ 
&	\nu_n \simeq 0.1 q_0^2 \rho_i^2 \nu_{ii}. \label{nundef}
\end{align}
The formulae for
the neoclassical diffusivity
and viscosity
in  the large-aspect-ratio,
concentric-circular-flux-surface,
banana (\(\nu_{ii} \ll v_{thi}/Rq\)) regime,
were taken from \cite{hintonwong1985nit}. 
In \cite{hintonwong1985nit}, the ion-ion collision frequency is defined as:
\begin{equation}
	\nu_{ii} = \frac{\sqrt{2 \pi}  Z_i^4 n_i e^4 \ln \Lambda}{T_i^{3/2} m_i^{1/2}}
	\label{nuiidef}
\end{equation}
(where $\ln \Lambda$ is the Coulomb logarithm). 
Thus, the neoclassical transport is a function
of the temperature and density,
and scales with these quantities 
in a different way to the turbulent transport.
To determine the ratio
between the neoclassical and turbulent transport
it is necessary to choose
specific values for the temperature and density.
Here, as in \cite{highcock2010transport}, 
we take
$\nu_{ii}=\nu_{ii}^N/2$.
Later, when we consider the case of specific 
tokamaks, we will use typical values of $n_i$ and $T_i$ 
to estimate $\nu_{ii}$.

Using the results \eqref{chindef} and \eqref{nundef}
the neoclassical Prandtl number
can be calculated from the parameters given in
\eqref{params}
and is found to be
\begin{equation}
	\mathrm{Pr}_n = \frac{\nu_n}{\chi_n} \simeq 0.01 \ll \prt.
	\label{prndef}
\end{equation}
Thus the neoclassical Prandtl number is much smaller than the turbulent Prandtl number (see Fig. \ref{fluxes}(c)).
In other words, the neoclassical transport of momentum is much smaller than the neoclassical transport of heat,
in contrast to turbulent transport,
for which the fluxes of momentum and heat are comparable.
While the formulae \eqref{nundef} and \eqref{chindef} are approximate,
we emphasise that the qualitative result
of the following section
is not affected by small changes in the values of \(\chi_n\) and \(\nu_n\),
provided that the property $\mathrm{Pr}_n \ll \mathrm{Pr}_t$ continues to hold.
In particular, this means that this qualitative result
is not affected by small changes in the value of $\nu_{ii}$.

\subsection{Bifurcation}
\label{transition}
\myfig{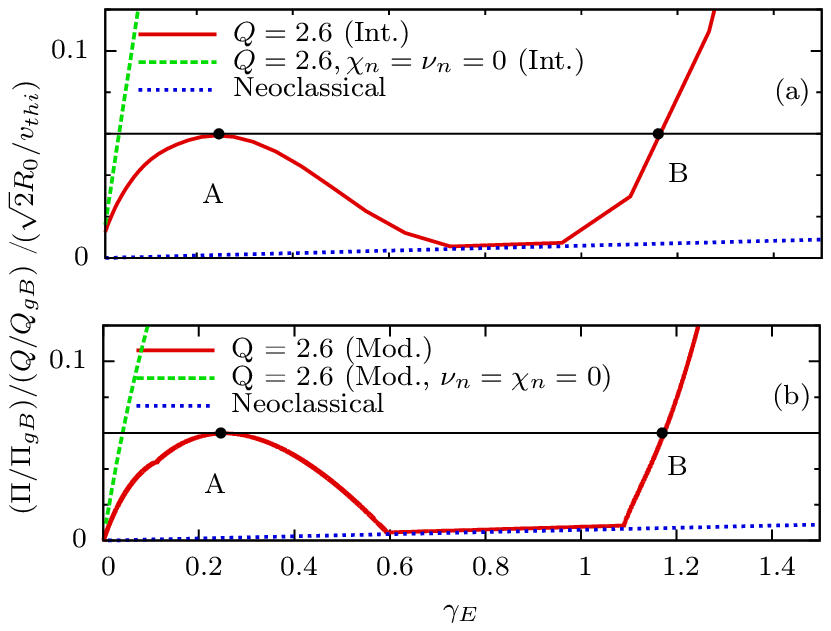}{Momentum flux divided by the heat flux (each normalised by the corresponding gyro-Bohm estimate) vs the flow gradient at a constant value of the heat flux $Q=2.6Q_{gB}$ plotted using (a) interpolation from the data explained in Section \ref{interpolationsec} and (b) the parameterisation from Section \ref{qbehaviour}. Also shown are the neoclassical contribution to the momentum flux (dotted line) and the momentum flux at constant heat flux \emph{without} the neoclassical contribution (dashed line). }

In Fig. \ref{transitionpg}(a), the momentum flux is once more plotted against $\fsr$,
but this time \emph{keeping the heat flux constant} at $Q/Q_{gB}=2.6$. 
The local maximum in the momentum flux still exists. 
Starting at this local maximum (point A), 
if the torque on the plasma was to be increased at constant $Q$, 
the flux of momentum could only increase 
if the flow shear were to jump to a much higher value (point B) 
where the PVG instability would drive turbulent momentum flux.
A bifurcation is manifest.

The inclusion of the neoclassical fluxes
is critical in obtaining this result. 
To illustrate this, \figref{transitionpg} also shows the momentum flux at constant heat flux
\emph{without} the neoclassical contribution to the fluxes, i.e., with $\chi_n=\nu_n=0$. 
If, in such a synthetic ``purely turbulent'' transport case,
the flow shear is increased at constant \(Q\),
the temperature gradient increases to maintain the turbulent heat flux,
and, as shown in \figref{transitionpg}, $\Pi/Q$ increases monotonically with \(\gamma_E\). 
The key property of neoclassical transport
that helps change this behaviour and produce a bifurcation
is that the neoclassical Prandtl number is much smaller than the turbulent Prandtl number. 
This means that as turbulent transport is suppressed,
the system goes through a regime
where the neoclassical contribution to the heat flux is significant,
while the neoclassical contribution to the momentum flux is not.
So as \(\Pi/Q\) is increased at constant \(Q\), the turbulence is suppressed, and the temperature gradient starts to rise, but as this happens more of the heat flux is transported neoclassically, and so the feedback loop breaks down: it is no longer necessary to increase the levels of turbulence to maintain the same \(Q\). 
The same is not true of the momentum: the neoclassical viscosity cannot make up for the lack of turbulence, and so the transport of momentum peaks and then falls.

At high flow shears the turbulent momentum flux increases rapidly once again; 
this turbulent flux is driven by the PVG,
as discussed in Section \ref{fluxessec}.
As a result, 
we observe that the bifurcation results in a turbulent state at B,
with a significant turbulent momentum flux. 
This is a substantial difference from the situation envisioned in \cite{waltz1997gyro},
where a transport bifurcation was predicted using a reduced quasilinear model.
Their model did not 
(and, being a quasilinear model, could not)
predict the existence
of the PVG-driven subcritical turbulence at high flow shears;
instead it predicted a bifurcation resulting in a non-turbulent state
where all transport was neoclassical. 
Thus a full nonlinear analysis is necessary
to describe accurately the reduced transport state produced by the bifurcation
that we find in a turbulent plasma.

Restoring dimensions,
we find that the bifurcation results
in a jump in the flow shear
from $0.25\,v_{thi}/R_0$
to $1.17\,v_{thi}/R_0$.
It is instructive to consider whether such values of shear 
appear in current devices.
Taking the measured values of the perpendicular flow shear, 
$T_i$ and $R_0$, in high-performance discharges in the JET and MAST tokamaks \cite{vries2009internal,field2004core},
we estimate that flow shears of up to approximately $v_{thi}/{R_0}$ have been recorded in both devices, and thus that values of shear of the order examined in this paper have been observed.

\section{Parameterised Model}
\label{qbehaviour}
Through simple interpolation, without making any assumptions about the way the fluxes depend on the gradients, we have shown the existence of a bifurcation in the gradients at constant fluxes. 
However, even interpolating one line of constant $Q$ requires a very large number of data points in the region of the line. 
To produce the required data set for \figref{transitionpg}(a),
we had to perform about 350 nonlinear gyrokinetic simulations, each run to saturation.
In order to extend our understanding of the bifurcation,
and of the range of flux values for which similar bifurcations can occur,
we will first consider the behaviour of the turbulent fluxes in further detail
and then use this analysis to construct
a simple parameterised model of the turbulent fluxes $Q_t$ and $\Pi_t$
as functions of $\fsr$ and $\tprim$. 

\subsection{Modelling $Q_t$}
\label{modellingqtsec}

\subsubsection{Dependence of $Q_t$ on $\tprim$}

\myfig{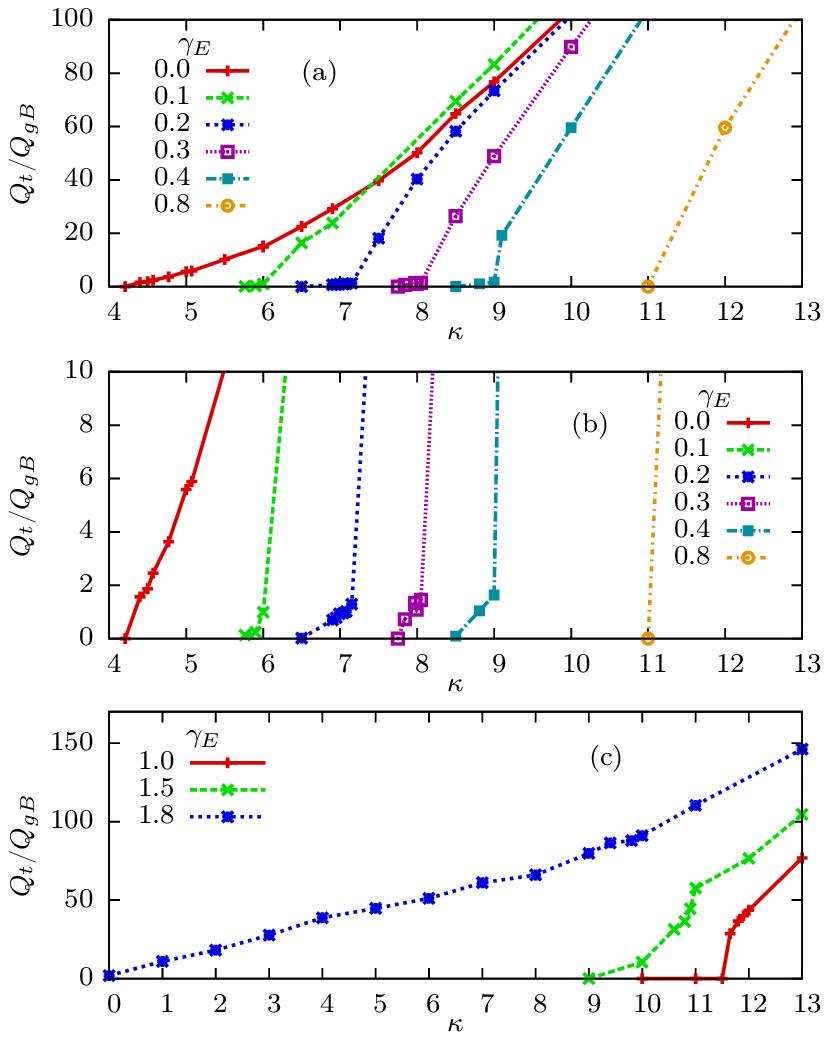}{ Turbulent heat flux vs the temperature gradient for different values of the flow shear, showing (a) low-shear values $\fsr<1\gstwofsrnorm$, (b) a close up of the low $Q_t$ region in (a), and (c) high-shear values $\fsr\geq1\gstwofsrnorm$}

We wish to construct a parameterised model of $Q_t$ as a function of $\fsr$ and $\tprim$.
In Section \ref{fluxessec} we described the dependence of $Q_t$ on $\fsr$;
we now consider the dependence of $Q_t$ on $\tprim$. 
Both experimentally \cite{manticastiffness} and numerically \cite{dimits:969},
it is usually found that $Q_t$ 
increases very sharply with $\tprim$ ---
a property known as stiff transport.
A recent experimental study \cite{manticastiffness} has
indicated that increasing the flow shear at low magnetic shear
might have the effect of reducing
the stiffness.
Fig. \ref{qvstprimbehaviour} shows that the effects of flow shear
on $\partial Q_t/\partial \tprim$ are in fact quite complex.
Let us consider the cases of low flow shear, $\fsr<1\gstwofsrnorm$, 
and high flow shear, $\fsr\gtrsim1\gstwofsrnorm$, separately.

For the case of $\fsr<1\gstwofsrnorm$, shown in \figref{qvstprimbehaviour}(a), the threshold in $\tprim$ above which turbulence can be sustained nonlinearly increases rapidly with $\fsr$, as the perpendicular shear suppresses the ITG instability. 
Above this threshold there are broadly three regions: low, intermediate and high $Q_t$.

At high $Q_t$, far above the threshold, 
for any value of flow shear
the heat flux eventually asymptotes to
the same dependence as it has at $\fsr=0$.
In other words, as the ITG drive becomes very strong,
the effect of flow shear becomes negligible.

At intermediate $Q_t$,
the heat flux rises rapidly from low values
to join the universal, high-$Q_t$ asymptotic. 
As $\fsr$ increases, because the threshold rises,
the heat flux rises more rapidly with $\tprim$ above the threshold;
thus at intermediate values of $Q_t$,
the stiffness $dQ_t/d\tprim$ increases with $\fsr$;
this is shown in \figref{stiffnessmodel}. 

At low $Q_t$, just above the threshold,
the heat flux rises very slowly,
that is, the stiffness $dQ_t/d\tprim$ is very low (see \figref{stiffnessmodel}).
This low $Q_t$ region is shown in detail in \figref{qvstprimbehaviour}(b).
Such is the sharpness of the transition
from the low-stiffness low-$Q_t$ region
to the high-stiffness intermediate-$Q_t$ region
that there appear to be two distinct thresholds.
The first threshold is the transition from no turbulence 
to non-zero turbulent transport.
Above the first threshold turbulence is present
but $Q_t$ rises slowly with $\tprim$ (the low-$Q_t$ region).
Above the second threshold $Q_t$ rises rapidly (the intermediate-$Q_t$ region).
These thresholds are plotted in \figref{modelq0}.
The low-stiffness region only exists for $0<\fsr<0.8\gstwofsrnorm$:
between $0.4<\fsr<0.8\gstwofsrnorm$ the first threshold joins the second
and the low-stiffness region disappears. 

At high flow shear, $\fsr\geq1$,
there are two principal differences to the case with low flow shear. 
Firstly, the low-stiffness region
is not present, and there is only one threshold.
Secondly,
the critical temperature gradient for turbulent heat transport starts to decrease
as the PVG drive re-enforces the ITG drive. 
When $\fsr=1.8$, the PVG drive is strong enough to drive turbulence at very low $\tprim$:
the threshold drops to zero.
As the threshold drops,
the transport stiffness at intermediate values of $Q_t$ decreases,
in a mirror image of the case at low flow shear.
At high $Q_t$, the heat flux still asymptotes to the universal $\fsr=0$ curve.

Thus, increasing $\fsr$ can both increase and decrease the nonlinear thresholds, 
and both increase and decrease the stiffness.
The rise and fall both in the thresholds
and the stiffness
can also be seen in
the finite-magnetic-shear simulations of 
\cite{barnes2011turbulent}. 
The principal differences between the zero-magnetic-shear case considered here 
and that case 
are that we have a higher value of the critical gradients for all $\fsr$,
and that we find a low-stiffness region at low values of $Q_t$ ---
a feature that seems to be absent at finite magnetic shear.

\myfig{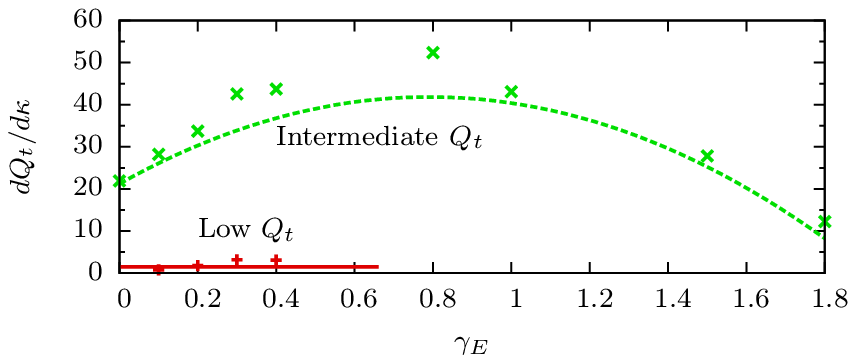}{The dependence of the heat transport stiffness $dQ/d\kappa$ on the flow shear in the low-$Q_t$ and intermediate-$Q_t$ regions, measured using simulations close to both thresholds (points, see Fig. \ref{qvstprimbehaviour}) and using the parameterised model of Section \ref{qtparameterisationsec} (lines).}
\myfig{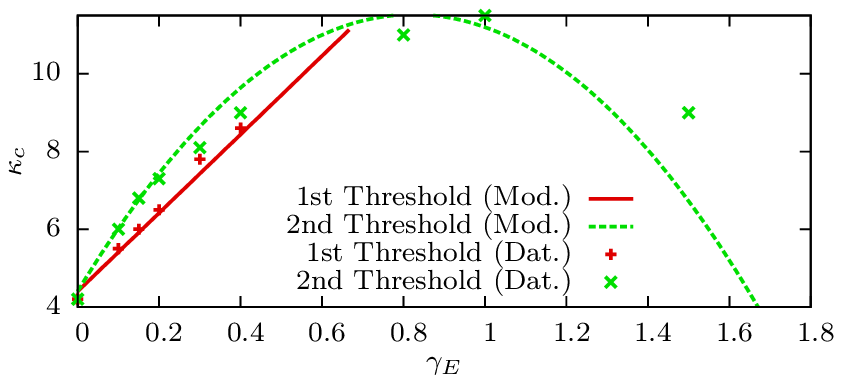}{The first and second critical thresholds, $\tprim_{c1}$ and $\tprim_{c2}$, measured using simulations close to both thresholds (points, see Fig. \ref{qvstprimbehaviour}), and using the parameterised model of Section \ref{qtparameterisationsec} (lines).}

\subsubsection{Parameterisation of $Q_t$}
\label{qtparameterisationsec}
In \cite{parra2011momentum} a simple model
was used to characterise the behaviour of the turbulent heat flux,
and describes the qualitative behaviour of the heat flux reasonably well. 
Here, we describe a more complex model 
that reproduces most of the features described above
and is \emph{quantitatively} close to the interpolated fluxes.
We consider only
the low and intermediate $Q_t$ regions,
as the bifurcation occurs
near the boundary between these regions.
In order to describe $Q_t$ in these regions,
we have to parameterise the two thresholds and the transport stiffness $dQ_t/d\kappa$.

We parameterise the first and second critical thresholds
as linear and quadratic functions of $\fsr$ respectively:

\begin{align}
&	\tprim_{c1} = \tprim_0 + \alpha_1 \fsr,\\ 
&	\tprim_{c2} = \tprim_0 + \alpha_2 \fsr + \alpha_3 \fsr^2, 
\end{align}
where $\tprim_0=4.4$ is the nonlinear threshold for turbulence at $\fsr=0$ and the parameters $\alpha_1=10.1$, $\alpha_2=17.4$ and $\alpha_3=-17.0$ are chosen to fit the data.
The modelled thresholds are plotted next to the measured thresholds in Fig. \ref{modelq0}.
As with the observed thresholds,
the first threshold joins the second between \(0.4<\fsr<0.8\gstwofsrnorm\). 

Next we parameterise the transport stiffness $dQ_t/d\kappa$.
The measured values of $dQ_t/d\kappa$ in the first and second regions are shown in \figref{stiffnessmodel}.
Between the first and second thresholds (the low $Q_t$ region),
$d(Q_t/Q_{gB})/d\kappa$ is modelled as a constant, $\alpha_4$.
In the intermediate $Q_t$ region, remembering the observation that
$dQ_t/d\kappa$ broadly rises and falls with the nonlinear thresholds, we 
allow the stiffness to depend on $\tprim_{c2}$.
Thus,
\begin{align}
	&\oov{Q_{gB}}\dbd{Q_t}{\tprim} = \alpha_4 && \tprim_{c1} < \tprim < \tprim_{c2},\\
	&\oov{Q_{gB}}\dbd{Q_t}{\tprim} = \alpha_5 + \alpha_6 \kappa_{c2} && \tprim > \tprim_{c2}.
	\label{secondregionstiffness}
\end{align} 
The parameters are $\alpha_4=1.5$, $\alpha_5=3.0$ and $\alpha_6=8.0$.
The model of $dQ_t/d\tprim$ is shown in \figref{stiffnessmodel}.

Thus $Q_t$ is parameterised as a piecewise linear function of $\tprim$.
It is zero below the first threshold,
has gradient $\alpha_4$ above the first threshold
and gradient $\alpha_5 + \alpha_6 \kappa_{c2}$ above the second:
\begin{align}
	\frac{Q_t}{Q_{gB}} &= 0 && \tprim < \tprim_{c1}, \nonumber\\
	\frac{Q_t}{Q_{gB}} &= \alpha_4 \lp \tprim - \tprim_{c1}\rp & &\tprim_{c1} < \tprim < \tprim_{c2},\nonumber\\
	\frac{Q_t}{Q_{gB}} &= \alpha_4 \lp \tprim_{c2} - \tprim_{c1} \rp  &&\label{qtmodel}\\ 
 &\qquad +\lp\alpha_5 + \alpha_6 \kappa_{c2}\rp  \lp \tprim - \tprim_{c2}\rp 	&&  \tprim > \tprim_{c2} . \nonumber
\end{align}
The model is compared with the data points in Fig. \ref{qvstprimmod}. 
To reproduce the bifurcation of Section \ref{transition} in a quantitatively correct manner,
it is in fact sufficient to represent accurately the two thresholds and the low stiffness region.
However, by also parameterising the variation of $dQ/d\kappa$,
we have provided a model which uses six parameters to describe the (complicated) behaviour of the heat flux over a wider parameter regime with reasonable accuracy.

\myfig{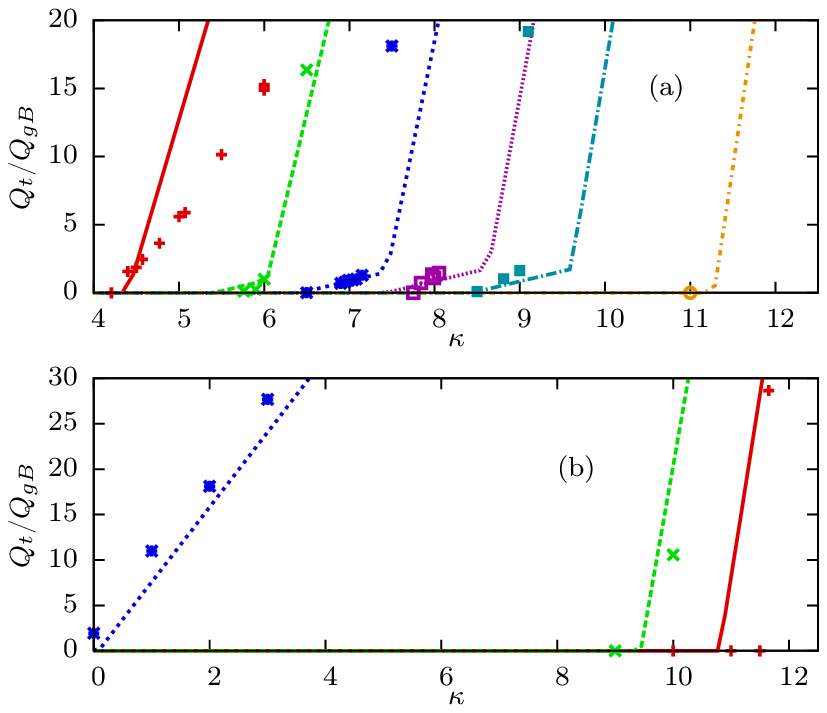}{The modelled heat flux (lines) shown along with the simulated data points for (a) low flow shear and (b) high flow shear. Legends as in Figures \ref{qvstprimbehaviour}(a) and \ref{qvstprimbehaviour}(c).}

\subsection{Modelling $\Pi_t$}

While the turbulent Prandtl number does vary with $\fsr$ and $\tprim$,
as discussed in Section \ref{prtsec},
this variation is relatively weak.
Thus we model $\Pi_t$ by assuming a constant turbulent Prandtl number:

\begin{align}
	\label{pitmodel}
	\frac{\Pi_t}{\Pi_{gB}} = \frac{Q_t}{Q_{gB}} \frac{\fsr}{\tprim}\frac{\sqrt{2} R_0 q_0}{r_0}\prt,\qquad
	\prt = 0.55.
\end{align}
The same choice was made in \cite{parra2011momentum}.

\subsection{The Modelled Bifurcation}

In Fig. \ref{transitionpg}(b) the model is used to replot 
the angular momentum flux at constant $Q/Q_{gB}=2.6$.
It has the same shape as the interpolated curve in \figref{transitionpg}(a),
and is quantitatively very close to it
at low flow shear ($\fsr < 1.0\gstwofsrnorm$). 
The agreement at higher flow shears is less good,
which reflects the fact that the second critical gradient is not really a quadratic.
Nonetheless, we consider this degree of precision adequate.
The model is used below to explore further properties of the transition
without the need for extremely large numbers of additional simulations.

\section{The Reduced Transport State}
\label{furtherobservationssec}

In Section \ref{transitionsec}, a transition was described leading to a reduced transport state. 
Here we use the parametric model that was designed in Section \ref{qbehaviour} to describe the properties of this state.

\subsection{Heat Flux at Constant $\Pi/Q$}
\myfig{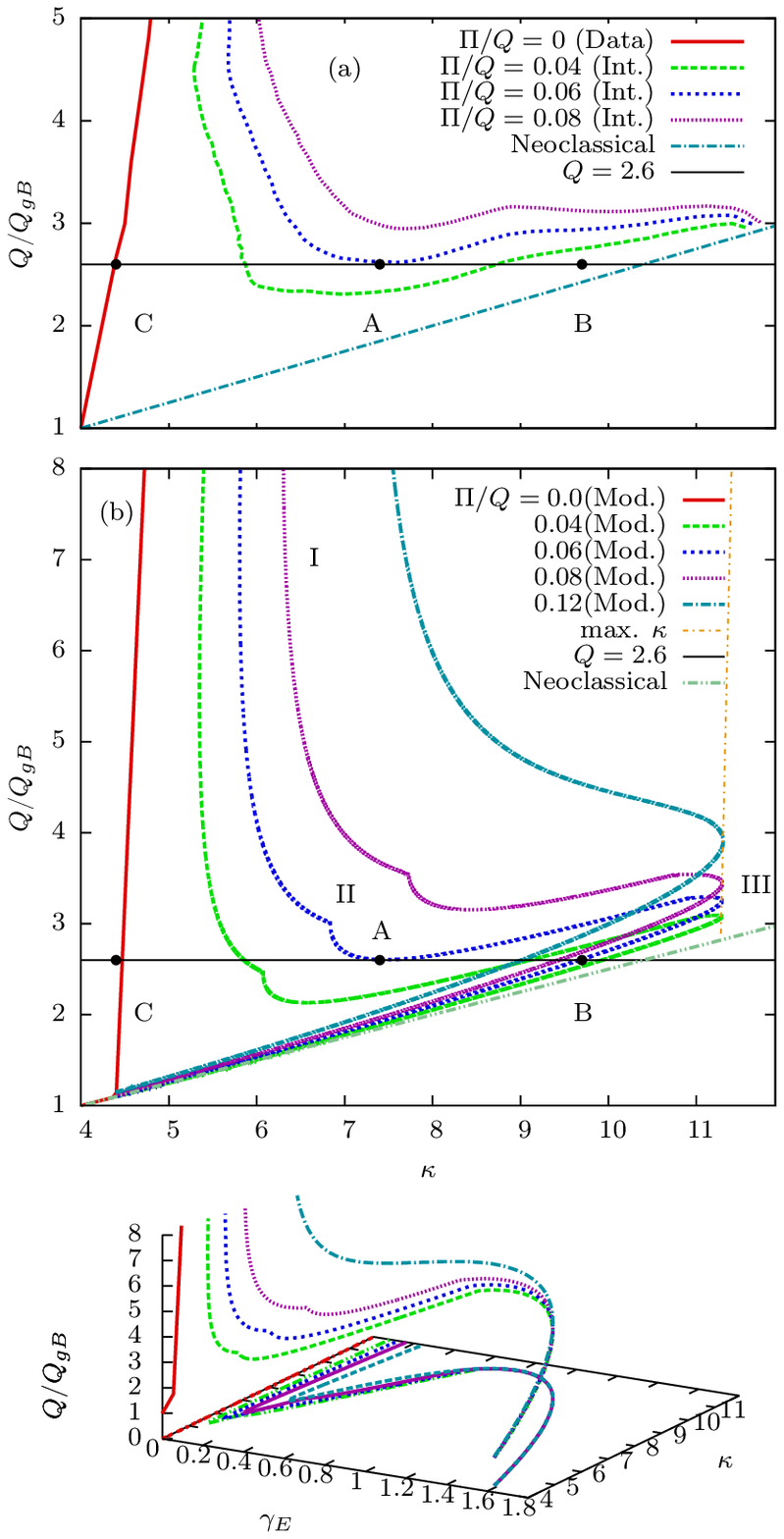}{Heat flux $Q$ vs the temperature gradient $\tprim$ at a constant ratio of the momentum flux to the heat flux $\Pi/Q$ (in units of $\sqrt{2} R_0 / v_{thi}$) plotted using (a) interpolation from the data (Sec \ref{interpolationsec}) and (b) the parameterised model (Sec \ref{qbehaviour}). Also shown is the neoclassical contribution to the heat flux. Interpolation in (a) is impossible near this neoclassical line, where the contours are closely spaced and \(\Pi/Q\) is multivalued. (b) also shows the maximum possible temperature gradient at low $Q$ for a given \(\Pi/Q\) (labelled ``max $\tprim$''). Fig. (c) plots the same curves as (b), showing both $Q$ against $\fsr$ and $\tprim$, and the curves projected on the $\fsr$-$\tprim$ plane, illustrating the increase in the flow shear along each curve of constant $\Pi/Q$. Points A and B in (a) correspond to points A and B in Fig. \ref{transitionpg}(a).}

\figref{transitionpg} showed the effect of varying $\Pi/Q$ whilst keeping $Q$ constant. 
In contrast, \figref{transitionqt} demonstrates the effect of varying $Q$ whilst keeping $\Pi/Q$ constant.
At high $Q$ and low temperature gradient (marked (I) in Fig. \ref{transitionqt}(b)),
increasing the heat flux has the effect of increasing the temperature gradient as might be expected:
this is the ordinary turbulent regime.
However, as $ Q/Q_{gB}$ is lowered below $\sim5$
there arises a counterintuitive situation where
decreasing $Q$ has the effect of raising the temperature gradient.
This is of course the combined effect
of the neoclassical Prandtl number
and the flow shear as described in Section \ref{transitionsec},
and also discussed in detail in \cite{parra2011momentum}.
In this region (marked (II) in \figref{transitionqt}(b)),
\(Q_t\) becomes comparable to \(Q_n\),
but as the heat flux is lowered,
the system cannot drop to a neoclassical state
because of the smallness of the neoclassical transport of momentum. 
Thus the constancy of the ratio $\Pi/Q$
(i.e., the large finite flux of momentum)
keeps the system in a turbulent state;
the temperature gradient and the flow shear rise,
and the curve representing $Q$ vs $\tprim$
becomes flatter and curls up
in order to maintain its distance from the neoclassical line. 
At large temperature gradients and flow shears
(region (III) in \figref{transitionqt}(b)),
the momentum flux increases rapidly due to the PVG drive
and the high flow gradient (see Fig \ref{fluxes}(b))
and so the curve rolls over and resumes its original downward trend. 
Eventually the heat flux asymptotes to the neoclassical value.

The last part of this trend, while physically obvious,
can only be seen using the modelled heat flux (Fig \ref{transitionqt}(b)),
as it is not feasible to interpolate
in the region near the neoclassical line
where the contours of constant $\Pi/Q$
are very closely spaced and \(\Pi/Q\) becomes multivalued. 

\subsection{Temperature Gradient Jump}

Interpolation cannot yield the temperature gradient after the transition directly:
as explained above, the contours of constant $\Pi/Q$ become too closely spaced
as they approach the neoclassical line in \figref{transitionqt}(a). 
However, the temperature gradient of the final state,
labelled B on both \figref{transitionpg} and \figref{transitionqt}(a,b),
can be calculated indirectly by using the value of $\fsr$ from \figref{transitionpg}(a)
and rearranging \eqref{prtdef} and \eqref{totalqp}
to give $\tprim$ as a function of $\fsr$, $\prt$, $\Pi/\Pi_{gB}$ and $Q/Q_{gB}$:

\begin{equation}
	\tprim = \frac{\prt \lp \sqrt{2} q_0 R_0 / r_0 \rp \fsr Q/Q_{gB}}
	{\Pi/\Pi_{gB} + \lp 4 R_0^2 q_0/v_{thi} r_0  \rho_i^2 \rp \fsr \lp \chi_n - \nu_n \rp }.
	\label{kappaform}
\end{equation}
Taking $\fsr=1.17$, $\prt=0.55$ yields a temperature gradient at point B of $\tprim=9.7$.
The temperature gradient at point A (the other side of the jump) is 7.4.
Alternatively, using our model,
the values of the temperature gradient can be read from \figref{transitionqt}(b) 
which results in an identical jump of $\tprim=7.4$ to $\tprim= 9.7$.
If we compare with the case of zero momentum input
and zero flow shear (point C on \figref{transitionqt}(a)),
we find that flow shear has enabled a total increase
in the temperature gradient of a factor \(9.7/4.5 = 2.2\).
If this were reproduced across the whole device, it could increase the core temperature by a factor of around 10, but this would require a more detailed 1D model (see \cite{barnes2010shear}), rather than the 0D model presented here.  
Note that while the jump in temperature gradient
caused by the bifurcation
(the transition from A to B)
is a striking feature,
a comparable contribution
to the increase of the temperature gradient 
arises from the incremental suppression of the turbulent transport
by the velocity shear (the difference between A and C).

\subsection{Neoclassical Heat Flux, Turbulent Momentum Flux}

It was noted earlier that the reduced transport state produced by the bifurcation is still turbulent,
with a momentum flux far greater than its neoclassical value (\figref{transitionpg}).
In \figref{transitionqt}(b), it can also be seen
that the reduced transport state does not lie exactly on the neoclassical line;
it does however lie very close to it,
which implies that the turbulent heat flux is much smaller than neoclassical.
Thus, the bifurcation takes the system into a state where
\emph{the heat is mostly transported neoclassically,
but the angular momentum is mostly transported by turbulence}.
The small ratio of $Q_t$ to $Q_n$
reflects the fact that the turbulence levels
in the reduced transport state are small.
The dominance of $\Pi_t$ over $\Pi_n$
given the low levels of turbulence
reflects the fact 
that the flow gradient is large
and the neoclassical momentum transport is very low
compared to the neoclassical heat transport
($\mathrm{Pr}_n \ll 1$).

\subsection{Bifurcation by lowering $Q/Q_{gB}$}

Starting from point A, if \(Q\) were to be reduced at constant \(\Pi/Q\),
the system would again be forced to jump to point B. 
In effect, what would happen is that the heat flux would become principally neoclassical;
the momentum flux would drop because $\mathrm{Pr}_n \ll \prt$, 
and a bifurcation would ensue in the manner described in Section \ref{transition}.
Thus, a decrease in the input heat could lead to a higher temperature gradient. 
We note, however, that \(Q\) is normalised to the gyro-Bohm value: \(Q_{gB} = \gstwoQnormflatbd\). 
Thus, decreasing \(Q/Q_{gB}\) could correspond to decreasing the deposition of heat,
but it could also correspond to an increase in temperature or density \cite{barnes2010shear}.

\subsection{Optimum $Q$}
\label{optimumqsec}
\figref{transitionqt}(b) shows that for every value of \(\Pi/Q\),
there is an optimum \(Q\) that gives a maximum temperature gradient
(the dotted line at $\tprim\sim11$ in Fig. \ref{transitionqt}(b)). 
The maximum temperature gradient increases with \(\Pi/Q\), but only slowly.  
The optimum value is the value of $Q$ such that:

\begin{equation}
	\left. \dbd{\kappa}{Q} \right|_{\Pi/Q} = 0
	\label{optqeq}
\end{equation}
and is plotted as a function of $\Pi/Q$ in \figref{transitionregion2}.

\subsection{Transition Region}

Finally, we will show that there is in fact only a limited range  of both \(Q\) and \(\Pi/Q\)
where bifurcations can happen in the way described above.
To illustrate,  we observe that no transition can occur along the contour
\(\Pi/Q=0.12 \sqrt{2} R_0 / v_{thi}\) in \figref{transitionqt}(b):
if $\Pi/Q$ is kept constant at this value,
decreasing \(Q/Q_{gB}\) from greater than its optimum value of $4.0$
increases the temperature gradient smoothly up to its  maximum value. 
The existence of a bounded region where such transitions can happen
is studied in detail by the authors of \cite{parra2011momentum},
who calculate, using a simple transport model,
the region in which transitions occur,
and derive a criterion necessary for their existence. 
We apply the analysis of \cite{parra2011momentum}
using the model for the turbulent fluxes given in \eqref{qtmodel} and \eqref{pitmodel},
to determine the range of values of $Q$ and $\Pi/Q$
for which bifurcations of the form we have described can occur.
In essence, bifurcations can only occur
when there exist multiple values of $\fsr$ and $\tprim$
that give rise to the same values of $Q$ and $\Pi/Q$. 
From \figref{transitionqt}(b),
it is clear that this is only possible for values of $\Pi/Q$
where there is a minimum in the curve of constant $\Pi/Q$,
and for values of $Q$ which lie between this minimum and 
either a maximum in the curve or 
the point where the curve intercepts the neoclassical line.
Thus the region in which transitions are possible is bounded by the curve

\begin{equation}
	\left. \pd{Q}{\tprim} \right|_{\Pi/Q} = 0,
	\label{transitioncondition}
\end{equation}
and by the neoclassical line.
This region is plotted in \figref{transitionregion2}. 

In order to give a clearer meaning to this diagram,
we use equations (\ref{qexp} - \ref{pioqexp})  
and typical  properties of plasma devices
taken from the ITER Profile Database \cite{roach2008profiledb}
(and listed in Table \ref{typicalvals})
to replot the region in which transitions can happen
in terms of the input beam power and
beam particle energy,
$P_{\mathrm{NBI}}$ and $E_{\mathrm{NBI}}$.
To generate these plots,
we also calculate the collision frequency $\nu_{ii}$
self-consistently using \eqref{nuiidef}, 
to take account of the varying strength
of the neoclassical transport in each device.
The transition regions for each device
are displayed in \figref{pnbienbiregion},
along with a point indicating the 
typical values of
$P_{\mathrm{NBI}}$ and $E_{\mathrm{NBI}}$
in each device.
\figref{pnbienbiregion} shows that
in which transitions can happen 
lie within an order of magnitude 
of the typical values.
It should be stressed
that with the simplified magnetic geometry (Section \ref{modelsec}) and
model of neoclassical transport (Section \ref{neoclassicaltransportsec}) used in this study,
and with the many assumptions about the 
way the energy and momentum are deposited
in the plasma (Section \ref{invertingsec}),
closer agreement could not be expected.
In particular, the assumptions of Section \ref{invertingsec}
are likely to overestimate the applied torque
and hence overestimate the beam energy 
needed for a transition.

\myfig{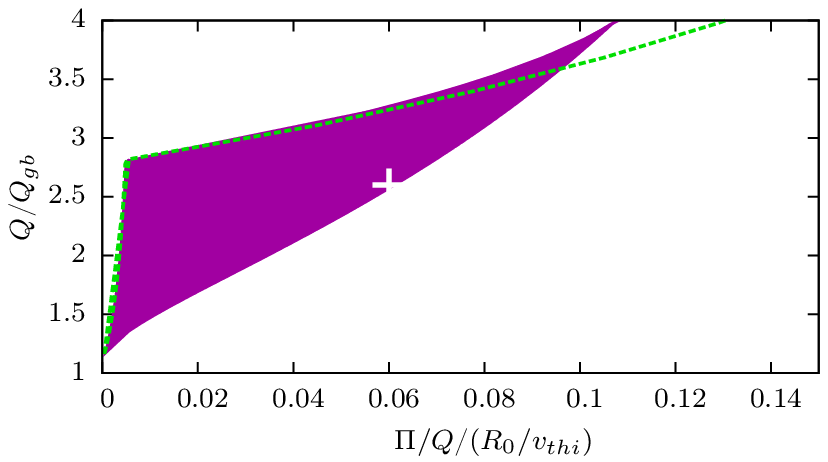}{The region in which bifurcations can occur, calculated using the analysis of \cite{parra2011momentum} and the parameterisation of the fluxes described in Section \ref{qbehaviour}. The cross indicates the location of the bifurcation described in Section \ref{transition}. The dashed line represents the optimum value of $Q$ for a given value of $\Pi/Q$, Eq. \eqref{optqeq}.}
\myfig{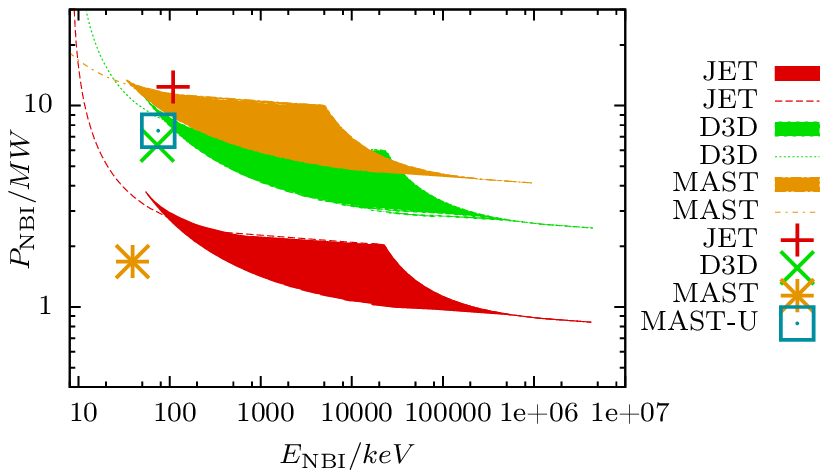}{The regions (shaded)
in which transitions can happen,
as a function of the beam power and the beam particle energy,
for three different devices.
The dashed lines indicate,
for each device,
the value of 
$P_{\mathrm{NBI}}$ for a given $E_{\mathrm{NBI}}$
which would lead to the optimum temperature gradient,
as described in Section \ref{optimumqsec}.
The points indicate typical values of 
$P_{\mathrm{NBI}}$ and $E_{\mathrm{NBI}}$
for each device.
The projected $P_{\mathrm{NBI}}$ and $E_{\mathrm{NBI}}$
for MAST Upgrade were taken from Ref. \cite{ccfewebsite}.
}

\begin{table}
\caption{Typical plasma properties from the ITER Profile Database \cite{roach2008profiledb}. The symbol $a$ denotes the minor radius of the device. The temperature was calculated as the mean of the core (TI0) and edge (TI95) temperatures.} 
\begin{center}
	\begin{tabular}{ l  r   r  r  r  r  r  r   }
		    \hline
		    \hline
				\bf		& $T_i$ ($eV$) & $n_i$ ($m^{-3}$) & $R$ ($m$) & $B_T$ ($T$) & $a$ ($m$)   \\
			 \hline
DIII-D & 2.7e+03 & 9.0e+19& 1.7e+00& 1.7e+00& 6.1e-01 \\
JET & 2.6e+03& 6.3e+19& 2.9e+00& 2.4e+00& 9.3e-01\\
MAST & 0.6e+03& 3.9e+19& 8.0e-01& 5.3e-01& 5.6e-01\\
			 \hline
			 \hline
	\end{tabular}
\end{center}
\label{typicalvals}
\end{table}

\section{Conclusions}
\label{conclusions}
We have determined the way the turbulent transport of heat
and toroidal angular momentum
is affected by the equilibrium gradients of temperature and velocity
over a wide range of those two parameters,
in the case of a plasma with a sonic background flow\footnote{
We have neglected the effect of the Coriolis force due to such a flow in our study. 
However, previous work \cite{peeters2007toroidal} indicates that its effect is to reduce the effective turbulent momentum diffusivity through a momentum pinch. 
This will make higher-flow-shear regimes easier to access but will not affect the bifurcation mechanism,
since the neoclassical transport of momentum remains much lower.
}.
We have extended the range of flow shears up to 20 times
the linear ITG instability growth rate 
calculated at zero flow shear. 
We have considered the zero-magnetic-shear regime.
We have used a low value of the safety factor, $q_0=1.4$,
which reduces the strength of the destabilising parallel velocity gradient drive
relative to the stabilising perpendicular velocity shear.
In so doing we have established the following key results:

\begin{enumerate}
	\item 
		The plasma is linearly stable to microinstabilities for all non-zero values of flow shear but subcritical turbulence can be sustained nonlinearly across much of this stable region.
	\item
		For velocity shear \(\fsr\lesssim 1\) 
		(we remind our readers that $\fsr$ is normalised to $v_{thi}/R_0$),
		increasing $\fsr$ 
		reduces the levels of turbulence
		driven by the ion temperature gradient.
		For \(\fsr\gtrsim 1\),
		the destabilising effect of the parallel velocity gradient
		becomes dominant and 
		increasing the flow shear increases the levels of turbulence and turbulent transport.
	\item
		For low flow shears, $\fsr\leq0.8$,
		and low heat flux, \(Q_t/Q_{gB}\lesssim2.5 \),
		flow shear reduces the thermal transport stiffness. 
		For low flow shears and higher $Q_t$, flow shear increases the thermal transport stiffness. 
		At high flow shears, $\fsr\gtrsim1$, the thermal stiffness is reduced as the PVG drive becomes dominant. 
	\item 
		For $\fsr\gtrsim0.1$
		the turbulent Prandtl number stays within the range 0.5--0.8
		across a wide range of temperature gradients.
		Thus, the turbulence transports heat and momentum with comparable vigour.

	\item
		For \(R_0/L_T\lesssim 11\),
		there is a large region around \(\fsr\sim1\)
		where the turbulence is completely quenched.
	\item
		Owing to the existence of this region,
		and the fact that the neoclassical Prandtl number
		is much lower than the turbulent Prandtl number,
		it is possible to trigger a bifurcation
		to a regime of high velocity and temperature gradients
		by either (i) increasing the momentum input at constant heat input
		or (ii) lowering \(Q/Q_{gB}\) at a constant ratio of heat to momentum input.
	\item
		In the high gradient (reduced transport) state
		that is reached via this bifurcation,
		the heat is principally transported neoclassically,
		whereas the momentum is principally transported by turbulence.
	\item
		For any given input of toroidal angular momentum, 
		there is an optimum input of of heat which maximises the temperature gradient. 
		Increasing the input of heat above this optimum
		can reduce the temperature gradient
		by increasing the levels of turbulence.
\end{enumerate}

\begin{acknowledgments}
We are grateful for helpful discussions with
I. Abel, P. de Vries, W. Dorland and G. W. Hammett.
This work was supported by
EPSRC (EGH, FIP), STFC (AAS), EURATOM/CCFE Association (MAB,CMR,SCC)
and the Leverhulme Network for Magnetised Plasma Turbulence.
The views and opinions expressed herein do not necessarily reflect those of the European Commission.
Computing time was provided by HPC-FF and by EPSRC grant EP/H002081/1.
Much of this work was carried out at
the Isaac Newton Institute for Mathematical Sciences, Cambridge, UK,
during the programme ``Gyrokinetics for Laboratory and Astrophysical Plasmas''. 
\end{acknowledgments}

\bibliography{references}
\end{document}